\documentclass[aps,prx,twocolumn,showpacs,amsmath,amssymb,floatfix,superscriptaddress]{revtex4-2}
\usepackage[outline]{contour}
\usepackage{color}
\usepackage{bbm}

\usepackage{graphicx}
\usepackage{dcolumn}
\usepackage[utf8]{inputenc}
\usepackage{graphicx}
\usepackage{epstopdf}
\usepackage{amsthm}
\usepackage{amsmath}
\usepackage{empheq}
\usepackage{eufrak}
\usepackage{bbm}
\usepackage{braket}
\usepackage{amssymb}
\usepackage{mathtools}
\usepackage[usenames,dvipsnames]{xcolor}
\usepackage{tikz}
\usepgflibrary{shapes.arrows}
\usetikzlibrary{calc, positioning, shapes.arrows}
\usepackage{cases}
\usepackage{smartdiagram}
\usepackage{latexsym}
\usepackage[colorlinks=true,citecolor=Cerulean,linkcolor=RubineRed,urlcolor=Cerulean]{hyperref}
\usepackage{cleveref}

\contourlength{0.5pt}
\contournumber{20}
\definecolor{S_Blue}{RGB}{0,135,252}
\definecolor{S_Red}{RGB}{214,13,63}
\definecolor{Blue}{RGB}{47,89,151}
\definecolor{S_Grey}{RGB}{150,150,158}
\definecolor{S_Yel}{RGB}{255,204,0}
\definecolor{S_Green}{RGB}{102,204,0}
\definecolor{S_Brown}{RGB}{154,41,41}
\DeclareFontFamily{OT1}{pzc}{}
\DeclareFontShape{OT1}{pzc}{m}{it}{<-> s * [1.15] pzcmi7t}{}
\DeclareMathAlphabet{\mathpzc}{OT1}{pzc}{m}{it}
\graphicspath{ {figures/figure1/}{figures/figure2/} }

\newcommand{\info}{\mathcal{I}_F} 
\newcommand{\dt}{\frac{d}{dt}} 
 
\newcommand{\cov}{\textnormal{cov}}
\newcommand{\LP}[1]{{\color{black}  #1}}

\begin{document}

\title{
Limits on the Evolutionary Rates of Biological Traits 
 }

\author{Luis Pedro Garc\'ia-Pintos}
\email{lpgp@lanl.gov}
\affiliation{Theoretical Division (T4), Los Alamos National Laboratory, Los Alamos, New Mexico 87545, USA}
\affiliation{Joint Center for Quantum Information and Computer Science and Joint Quantum Institute, NIST/University of Maryland, College Park, Maryland 20742, USA}

\begin{abstract}
 
This paper focuses on the maximum speed at which biological evolution can occur. I derive inequalities that limit the rate of evolutionary processes driven by natural selection, mutations, or genetic drift. These \emph{rate limits} link the variability in a population to evolutionary rates. In particular, high variances in the fitness of a population and of a quantitative trait allow for fast changes in the trait's average. In contrast, low variability makes a trait less susceptible to random changes due to genetic drift. The results in this article generalize Fisher's fundamental theorem of natural selection to dynamics that allow for mutations and genetic drift, via trade-off relations that constrain the evolutionary rates of arbitrary traits. The rate limits can be used to probe questions in various evolutionary biology and ecology settings. They apply, for instance, to trait dynamics within or across species or to the evolution of bacteria strains. They apply to any quantitative trait, e.g., from species' weights to the lengths of DNA strands. 

\end{abstract}

\maketitle

\section{Introduction}

Fisher's theorem of natural selection relates the rate of change in the average fitness of a population with the variability in fitness. It holds for evolutionary processes driven by natural selection~\cite{fisher1958genetical}. Fisher's result \LP{suggests} that variability serves as a resource by enabling fast evolution. However, the result is of rather limited validity: it does not apply to types that mutate or in the presence of genetic drift~\cite{price1972fisher,edwards2002fundamental,
e23111436}. 
\LP{Moreover, Fisher's focus was on the change in fitness. 
However, in countless instances biologists are interested in other quantitative attributes---or traits--- of individuals in a population.}
Here, I extend Fisher's results by studying (a) the rates of arbitrary biological traits and (b) general evolutionary processes that incorporate mutations and genetic drift.

This paper focuses on the rate $\frac{d\langle A \rangle}{dt}$ at which the average $\langle A \rangle$ of a quantitative trait $A$ changes. $A$ can represent any measurable trait in a population. For example, $A$ could be breeds' weights within a species, the maximum CO2 concentration at which different species can survive, or DNA lengths in bacteria strains. The results apply to a range of settings in evolutionary biology and ecology where one is interested in evolution of traits.

Traits' evolution rates have been studied extensively in quantitative biology. Reference~\cite{RateLimits2011} focuses on the biological factors that influence maximum growth rates. References~\cite{Hoffmann2005limits,hoffmann2014evolutionary} and~\cite{RateLimits2013,RateLimits2019} study how genetic variance and a population's structure affects trait evolution, respectively. There's also extensive data-based work on traits' evolutionary rates. As one example, Ref.~\cite{RateLimits2012} studies the maximum growth rates in mammals.

In the field of applied mathematics, extensions of Fisher's theorem have also been considered. Reference~\cite{Basener2018}, for example, revises Fisher's results by studying the effect of mutations on the change in the average fitness of a population (note, though, that it does not focus on other traits). Reference~\cite{lion2018theoretical} includes an interesting mathematical generalization of Fisher's theorem to arbitrary traits and dynamics beyond natural selection. References~\cite{marsland2019thermodynamic,ConstraintsSelectionPRR2021} rely on uncertainty relations from stochastic thermodynamics to study biological processes.

To derive limits on biological evolutionary rates, I will leverage techniques that have proven useful to study the maximum speed of physical processes. The most related results have appeared in Refs.~\cite{adachi2022universal,ItoUgh2023}. Both articles rely on  information theory to bound the evolutionary rates of arbitrary traits. The results in Ref.~\cite{adachi2022universal} hold for arbitrary processes. However, they do not discriminate how different evolutionary forces affect rates. In contrast, the results in this article isolate the contributions of natural selection, mutations, and genetic drift to evolutionary rates. The results in Ref.~\cite{ItoUgh2023} separate contributions from natural selection and mutations, but do not account for genetic drift. Moreover, the results in Refs.~\cite{adachi2022universal,ItoUgh2023} involve information-theoretic quantities (versions of the Fisher information) that can be hard to evaluate unless one possesses enough knowledge about the dynamics of the system. In contrast, the main results in this article depend on averages and standard deviations that are often more accessible from experimental data.

\begin{figure*}
  \centering   
 \includegraphics[trim=00 00 00 00,width=0.50\textwidth]{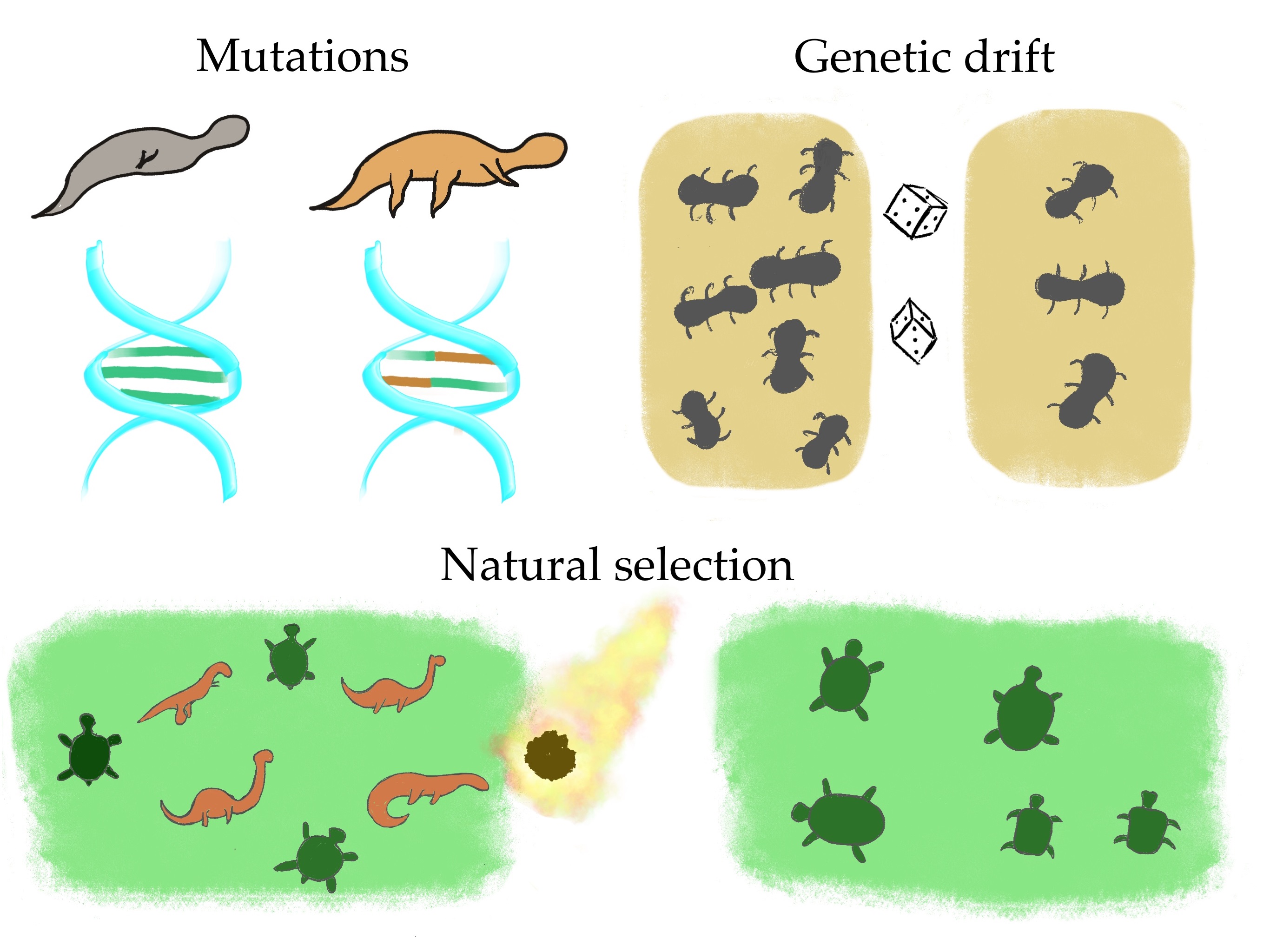} 
 \qquad
 \includegraphics[trim=00 00 00 00,width=0.42\textwidth]{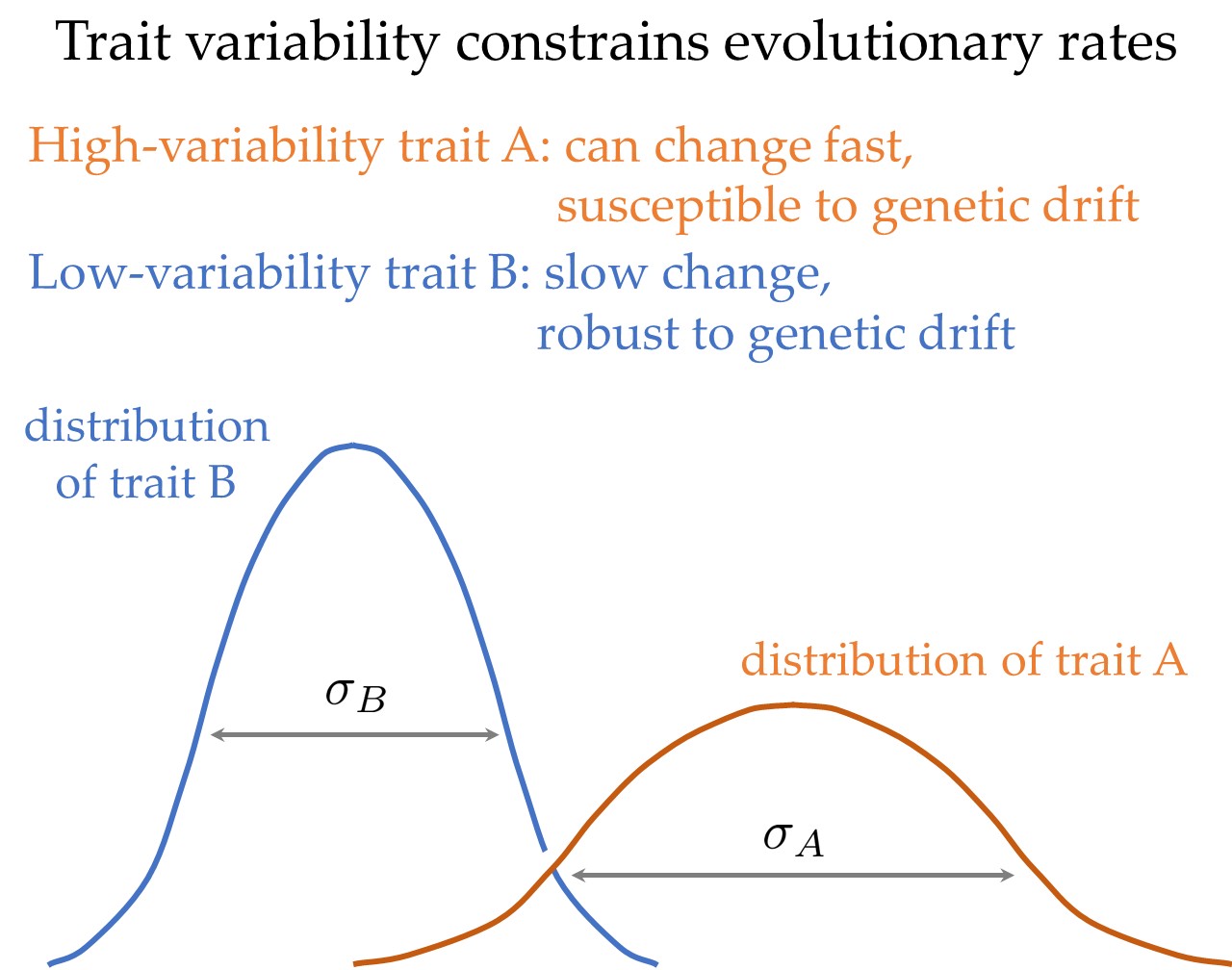}\\
 $\xrightarrow[]{\hspace{72pt} \textnormal{\emph{time}}\hspace{72pt}} \hspace{72pt}\hspace{72pt}$ \hspace{69pt}\hspace{42pt}
\caption{\textbf{Evolutionary rate limits.---} 
 In this work, I derive inequalities that bound the evolutionary rates of quantifiable traits. The rate limits apply to populations that evolve under mutations, natural selection, and random genetic drift. Mutations take a preponderant role in biology, enabling evolutionary changes that can lead to new types. Such new types can have \LP{different} fitness than their ancestors. Natural selection is the process by which the frequencies of fitter types tend to increase relative to their peers in a given environment \LP{(in the illustration, turtles became fitter to survive in a drastically-changed environment.)} A population's evolution can also be affected by chance, known as genetic drift \LP{(in the illustration, a random event reduced the population of an ant colony.)} The results in this Letter imply that, under natural selection, a trait $A$ with high variability $\sigma_A$ can evolve faster than a trait $B$ with small variability $\sigma_B$. At the same time, the trait with higher variability is more susceptible to the effect of random fluctuations in populations due to genetic drift. In this way, the rate limits can be used to discriminate quantifiable traits in terms of their maximum evolutionary rates and their responsiveness to different evolutionary forces.
  \label{fig:fig1}
}
\end{figure*}

The main outcome of this paper is a set of inequalities that constrain the evolution rate of  any quantitative biological trait $A$
(e.g., of a particular phenotype)
 in terms of simple properties of the system of interest. 
 Specifically, knowledge of expectation values and  variances of $A$ and of the fitness $f$ of a population suffices to evaluate the inequalities (see details in Secs.~\ref{sec:mutations} and~\ref{sec:stochastic}). 
 In this way, slowly evolving traits can be discriminated from those that can rapidly change without the need to exactly solve the complex dynamics of the system, as pictorially illustrated in Fig.~\ref{fig:fig1}. 

 In Section~\ref{sec:general}, I review general limits on traits' evolutionary rates that hold for any biological system. These results are very general but can sometimes be hard to evaluate. In Sections~\ref{sec:naturalselection},~\ref{sec:mutations}, and~\ref{sec:stochastic}, I derive rate limits that hold for systems driven by (a) natural selection, (b) natural selection and mutations, and (c) natural selection, mutations and stochastic forces (genetic drift), respectively. The results in Secs.~\ref{sec:naturalselection}-\ref{sec:stochastic} are tailored to evolutionary processes and  can be evaluated with knowledge of averages and standard deviations. I conclude in Sections~\ref{sec:experiments} and~\ref{sec:discussion} with potential connections to experiments and a discussion.

\vspace{12pt}
\noindent\fbox{ \parbox{0.46\textwidth}{
    \textbf{Non-technical summary.} How fast can biological evolution occur in nature?  What properties of a population enable drastic changes? To what extent are evolutionary rates affected by different driving forces such as natural selection, mutation, or random processes? I address these questions mathematically by deriving inequalities that limit the speed of evolutionary processes. The inequalities take the form of trade-off relations that relate the rate of change of a biological trait with its variance and with the variance in the fitness of a population. In this way, slowly evolving traits can be discriminated from those that can rapidly change. Evaluating the inequalities can be done without exactly solving the complex dynamics of the system. In broad terms, the results in this paper quantify the extent by which variability in a trait can lead to faster evolution.
    }}

\section{Rate limits on biological processes}
\label{sec:general}
Consider a set of types denoted by indexes $\{j\}$ with an evolving number of individuals $n_j = n_j(t)$ and a total (evolving) population $N = \sum_j n_j$ (I will also analyze results in terms of frequencies of types in the next section). 
Let $r$ denote the \emph{growth rate} of the population, with components $r_j \nobreak \coloneqq \nobreak \dot n_j/n_j$ for type $j$~\cite{Rice,vandermeer2010populations}.
Then, the rate of change in the expectation value $\langle A \rangle = \sum_j a_j n_j/N$ of a quantitative trait $A$ satisfies
\begin{align}
\label{eq:Price}
\frac{d \langle A \rangle}{dt} &= \sum_j \dot a_j \frac{n_j}{N}  + \sum_j a_j 
\frac{d}{dt}\frac{n_j}{N} = \langle \dot A \rangle + \cov(A,r).
\end{align} 
The covariance $\cov(A,B) \nobreak \coloneqq \nobreak \left\langle  A B \right\rangle \nobreak -  \nobreak \left\langle A \right\rangle \!\left\langle B \right\rangle$ characterizes the correlations between two quantities $A$ and $B$~\cite{Rice}. 
See the Supplementary Material for the proof of Eq.~\eqref{eq:Price}. Throughout this work, I use $d a /dt$ or $\dot a$ interchangeably to denote time derivatives.

In the context of evolutionary biology, Eq.~(\ref{eq:Price}) is known as the (time-continuous) Price equation~\cite{Rice,
price1970selection,frank2012Price,queller2017fundamental}. \LP{Equation~\eqref{eq:Price} is a mathematical identity that holds under very general assumptions (e.g., differentiability) about the underlying dynamics of the system.} Similar equations describe the dynamics of classical systems with evolving probability distributions~\cite{Nicholson_2020} and of open quantum systems~\cite{LPGP2022}. Reference~\cite{frank2020fundamental} discusses the connections between the Price equation in biology and statistical physics, while Ref.~\cite{lion2018theoretical} contains a detailed analysis of the way that different biological factors contribute to each term in the Price equation. 
 
 The first term in the Price equation~\eqref{eq:Price}, $\langle \dot A \rangle \nobreak=\nobreak \sum_j \dot a_j n_j/N$, describes changes in $\langle A \rangle$ due to explicit time dependence in the values $a_j$ of the trait. 
The second term in the Price equation, $ \sum_j  a_j \tfrac{d}{dt}(n_j/N) \nobreak =\nobreak  \cov(A,r)$, corresponds to the change in the average $\langle A \rangle$ of a trait due to the population changes. Using that the covariance is bounded by the product of standard deviations~\cite{Rice,colquhoun1971lectures}, the latter term is constrained by 
\begin{align}
\label{eq:speedlimit}
\left| \frac{d\langle A \rangle}{dt} - \langle \dot A \rangle \right|  = \big| \cov(A,r) \big| \leq \sigma_A \, \sigma_r.
\end{align}
Here, 
\begin{subequations}
\begin{align}
    \sigma_A^2 &\coloneqq \langle A^2 \rangle - \langle A \rangle^2 = \sum_j a_j^2 \frac{n_j}{N} - \bigg( \sum_j a_j \frac{n_j}{N} \bigg)^2, \\
    \sigma_r^2 &\coloneqq \langle r^2 \rangle - \langle r \rangle^2 = \sum_j r_j^2 \frac{n_j}{N} - \bigg( \sum_j r_j \frac{n_j}{N} \bigg)^2,
\end{align}
\end{subequations}
are the variances of the trait $A$ and of the growth rate $r$, respectively. (The standard deviation $\sigma_A$ is the square root of the variance.) 
    
The inequality in~Eq.~(\ref{eq:speedlimit}) sets a fundamental trade-off between the rate of change of a biological trait $A$ and (i) the variability $\sigma_A$ in the trait and (ii) the variability $\sigma_r$ in the growth rate $r$ with which the populations $\{n_j\}$ change: fast evolution requires a variable trait and variable population growth rate. I equate the variability of a trait $A$ with its standard deviation $\sigma_A =\sqrt{\langle A^2 \rangle - \langle A \rangle^2}$~\cite{colquhoun1971lectures}. Note that, from Eq.~(\ref{eq:Price}), one concludes that the sign of $d\langle A \rangle - \langle \dot A \rangle$ depends on whether the trait $A$ is positively or negatively correlated with the growth rate $r$. 

 Trade-off relations like Eq.~(\ref{eq:speedlimit}), typically referred to as speed limits in physics, also constrain the dynamics of quantum~\cite{mandelstam1991uncertainty,deffner2017quantum,LPGP2022} and classical~\cite{Nicholson_2020} physical systems. In the latter setting, the variance of $r$ is replaced by the Fisher information $\info$. The Fisher information is a measure of the speed with which a probability distribution evolves. For a time-dependent probability distribution $p_j$, it is given by $\info \coloneqq \sum_{j} p_j \! \big( \frac{\dot p_j}{p_j} \big)^2$~\cite{cramer1999mathematical}.  Defining $p_j \coloneqq n_j/N$ as the frequency of occurrence of type $j$, I prove in the Supplementary Material that indeed $\sigma_r = \sqrt{\info}$. Related limits to biological systems in terms of the Fisher information have been considered in Refs.~\cite{adachi2022universal,ItoUgh2023}. 

While the constraint in Eq.~(\ref{eq:speedlimit}) is extremely general, its practical usefulness may be hindered by the difficulty in relating the variability in the growth rate $\sigma_r$ (or, equivalently, the Fisher information) to the relevant parameters that govern the dynamics of a concrete system. My goal is to derive bounds (mathematical inequalities) on the rate of change of biological quantities ---or \emph{rate limits} for short--- tailored to evolutionary processes.

\section{Limits to replicator evolutionary processes}
\label{sec:naturalselection}
Under the assumption that mutation rates between types are negligible, the \emph{replicator equation},
\begin{align}
\label{eq:Replicator}
\dot p_j = p_j \big( f_j - \langle f \rangle \big),
\end{align}
can be used to model population dynamics~\cite{MAYNARDSMITH1974209,SCHUSTER1983533,Rice} (see Refs.~\cite{CrutchfieldPRE2003,ReplicatorEcon2010, ReplicatorPNAS2014, bloembergen2015evolutionary} for applications of the replicator equation to various other fields). Here, $p_j \coloneqq n_j/N$ is the frequency of occurrence of type $j$, and the \emph{fitness} $f_j \nobreak \equiv \nobreak f_j(\{p_k\},t)$ characterizes whether the frequency $p_j$ of a type increases or decreases: the populations of types with positive excess fitness, $f_j \geq \langle f \rangle$, tend to grow relative to their peers~\cite{smith1998evolutionary}.

Using Eq.~(\ref{eq:Replicator}), it holds that $r_j = \dot p_j/p_j + \dot N/N = f_j - \langle f \rangle + \dot N/N$, which in turn leads to $\sigma_r = \sigma_f$. Thus, Eq.~(\ref{eq:speedlimit}) implies that any evolutionary process that can be modeled by the replicator equation is constrained by 
\begin{align}
\label{eq:speedlimitReplicator}
\left| \frac{d\langle A \rangle}{dt} - \langle \dot A \rangle \right| &= \big| \cov(A,r) \big| = \big| \cov(A,f) \big| \leq \sigma_A \, \sigma_f, 
\end{align}
where I used that the covariance is invariant under the addition of uniform functions~\cite{colquhoun1971lectures}. The rate of change in natural selection processes is thus limited by the variability of the fitness of the population and the variability of the quantity of interest. 

The rate limit in Eq.~(\ref{eq:speedlimitReplicator}) for the replicator equation implies constraints on the dynamics of arbitrary quantitative traits $A$. If $A$ has no explicit time dependence, i.e., if the $a_j$'s are constant, then Eq.~(\ref{eq:speedlimitReplicator}) becomes a bound on the evolutionary rates $d \langle A \rangle/dt$, discriminating slowly evolving traits from those that can change rapidly. In plain terms, the inequality says that evolution is slow for systems with homogeneous fitness functions, where $\sigma_f \approx 0$ (neutral selection regime). In contrast, evolution can be faster on systems with a diverse population such that $\sigma_f $ is large (natural selection regime). 
 This mathematically formalizes and quantifies the common understanding that diversity serves as an evolutionary resource~\cite{naeem2012functions,Hoffmann2005limits,hoffmann2014evolutionary}, in this case by enabling fast evolution whenever a trait and fitness have variability across a population. These variabilities can only occur in sufficiently diverse populations.

 \LP{
Equation~\eqref{eq:speedlimitReplicator} involves terms that may be reminiscent of the breeder's equation. In the breeder's equation, $\Delta \langle A \rangle = S h^2$, the net
change $\Delta \langle A \rangle$ in a trait is governed by a measure of heritability ($h$) and the selection coefficient $S$. 
The selection coefficient measures covariance between fitness and a trait, so $S = \cov(A,f)$~\cite{heywood2005exact}. Then, Eq.~\eqref{eq:speedlimitReplicator} relates $S$ in the breeder's equation to the trait and fitness variabilities: $|S| \leq \sigma_A \, \sigma_f$. This may be useful in scenarios where the heritability is known but the selection coefficient is not, or, possibly, to study trait changes beyond the regime of applicability of the breeder's equation~\cite{morrissey2010danger}.
(Note, in particular, that the breeder's equation is less general than Price's~\cite{heywood2005exact}.)
}

 It is natural to wonder whether the left-hand and right-hand sides of the inequality~(\ref{eq:speedlimitReplicator}) are similar (i.e., whether the bound is saturated). When this happens, knowledge of the standard deviations $\sigma_A$ and $ \sigma_r$ suffices to estimate the evolutionary rate. 
This happens whenever $A$ has a linear relationship with the growth rate $r$, i.e., $a_j \nobreak\propto\nobreak r_j \nobreak+\nobreak c$ where $c$ is independent of $j$, in which case $\cov(A,r)\nobreak=\nobreak \sigma_A \, \sigma_r$~\cite{Nicholson_2020}. This is the case for the fitness function under replicator dynamics, so Eq.~(\ref{eq:Price}) yields
\begin{align}
\label{eq:thmNatSel}
 \frac{d\langle f \rangle}{dt} - \langle \dot f \rangle   =   \sum_j \dot p_j f_j  = \cov(f,f) =  \sigma_f^2.
\end{align}
This corollary of the general rate limit~\eqref{eq:speedlimitReplicator} provides a simple proof of Fisher's \emph{fundamental theorem of natural selection}~\cite{fisher1958genetical,price1972fisher,edwards2002fundamental,e23111436}. It shows that Fisher's claim is exact for  (i) evolutionary processes modeled by a replicator equation with (ii) fitness functions that are independent of time, in which case $  \frac{d \langle f \rangle}{dt}    = \sigma_f^2 $. In situations with more general fitness functions $f_j = f_j(\{p_k\},t)$, Eq.~(\ref{eq:thmNatSel}) provides a generalized version of Fisher's theorem whereby the velocity with which fitness changes due to changes in population frequencies equals the fitness variance.

\section{Limits to evolutionary processes with mutations} 
\label{sec:mutations}
Mutations are a crucial driving force in realistic evolutionary processes~\cite{gregory2009understanding,carlin2011mutations,hershberg2015mutation}. Mutations between types can be described by the replicator-mutator, or quasispecies model: 
\begin{align}
\label{eq:ReplicatorMutation}
\dot p_j =  \sum_{k } p_k  Q_{kj} f_ k - p_j \langle f \rangle.
\end{align}
 $Q_{kj} \nobreak\geq\nobreak 0$ is a dimensionless transition matrix that models mutations between types, which satisfies $\sum_{j} Q_{kj} \nobreak =\nobreak 1$~\cite{eigen1982stages, Hofbauer1985,wilke2005quasispecies, nowak2006evolutionary}. The replicator equation~(\ref{eq:Replicator}) is recovered when the mutation matrix is the identity, $Q_{kj} = \delta_{kj}$.

The general inequality~(\ref{eq:speedlimit}) holds in this case, too. Note, though, that while for the replicator dynamics the variance in the growth rate equals the variance in fitness, $\sigma_r^2 \nobreak = \nobreak \sigma_f^2$, this is no longer the case for dynamics with mutations. Relating the variance in the growth rate (or equivalently, the Fisher information) to biologically relevant quantities in concrete settings remains an interesting problem to be explored.

Alternatively, I define the mutation-driven distribution
\begin{align}
\label{eq:mutation-drivenProb}
\Pi_j \nobreak \coloneqq \nobreak \sum_k p_k Q_{kj}.
\end{align}
I interpret $\Pi$ as the frequency with which a given type would hypothetically occur in the future if evolution were only driven by mutations, or, possibly more biologically relevant, in regimes where strong mutation dominate over natural selection processes~\cite{gregory2009understanding}. Note that $\Pi = p$ in the mutation-less regime.
 
 Then, I prove in the Supplementary Material one of the main results of this work:
\begin{align}
\label{eq:speedlimitReplicatorMutation}
\left| \frac{d\langle A \rangle}{dt} - \langle \dot A \rangle - \langle f \rangle \Big( \langle A \rangle_\Pi - \langle A \rangle \Big)  \right| \leq \sigma_A^\Pi \, \sigma_f.
\end{align}
$\langle A \rangle_\Pi \coloneqq \sum_j \Pi_j a_j$ and $\sigma_A^\Pi \coloneqq \sqrt{\langle A^2 \rangle_\Pi - \langle A \rangle_\Pi^2}$ are the average and the standard deviation evaluated in the mutation-driven distribution $\Pi$. I emphasize that the general result in Eq.~(\ref{eq:speedlimit}) does not imply the rate limit~(\ref{eq:speedlimitReplicatorMutation}). The replicator-mutator evolutionary model, given by Eq.~(\ref{eq:ReplicatorMutation}), was crucial to derive the latter bound. 

The rate of change of any trait $A$ is thus constrained by 
the quantity's standard deviation evaluated in the mutation-driven distribution $\Pi$, and the standard deviation of the fitness of the system. As in the mutation-less setting, diversity in the population is seen to give rise to less constrained evolution rates.

One could be puzzled by the appearance of a term in Eq.~(\ref{eq:speedlimitReplicatorMutation}) that directly depends on the average fitness $\langle f \rangle$ of the population and not just on relative fitness values. After all, shifting the fitness by an additive constant $f_j \nobreak \rightarrow \nobreak  f_j \nobreak +\nobreak  c$ in the replicator equation~(\ref{eq:Replicator}) does not affect population dynamics. However, this is not the case in the mutator-replicator equation~(\ref{eq:ReplicatorMutation}), where an additive constant $c$ on the fitness function leads to a change in $\dot p_j$ of $c(\Pi_j \nobreak -\nobreak  p_j)$, which depends on how the distribution $p$ and mutation-driven distribution $\Pi$ differ. The absolute values of the fitness function play a dynamical role in a system with mutations, and this is manifested in the rate limit~(\ref{eq:speedlimitReplicatorMutation}).

\LP{To simplify interpretation}, let us momentarily consider the case when $A$ does not explicitly depend on time, i.e., the $a_j$'s are constant. Then, Eq.~(\ref{eq:speedlimitReplicatorMutation}) and the triangle and reverse triangle inequalities imply upper and lower rate limits, 
\begin{subequations}
\label{eq:upperlowerSL}
\begin{align}
\label{eq:lowerSL}
\left| \frac{d \langle A \rangle}{dt} \right|  &\geq  \langle f \rangle \Big|  \langle A \rangle_\Pi - \langle A \rangle  \Big|  - \sigma^\Pi_A \, \sigma_f, \\
\label{eq:upperSL}
\left| \frac{d \langle A \rangle}{dt} \right| 
&\leq   \sigma^\Pi_A \, \sigma_f +  \langle f \rangle \Big| \langle A \rangle_\Pi - \langle A \rangle \Big|.  
\end{align}
\end{subequations}
 Here, one can identify two distinct sources that contribute to the evolutionary rates of the system. One of the sources, $\sigma^\Pi_A \, \sigma_f$, involves the standard deviation of the fitness function, and the standard deviation of the trait of interest evaluated in the mutation-driven distribution. The remaining term depends on the averages of fitness and the quantity of interest. This is somewhat reminiscent of the speed limits for open quantum systems, where two distinct sources to the dynamics of a system lead to additive contributions to the ultimate speed with which a quantity can evolve, which in turn allows to derive lower bounds on speed~\cite{LPGP2022}. A different bound on evolutionary rates under natural selection and mutations was derived in Ref.~\cite{ItoUgh2023} in terms of contributions to the Fisher information (i.e., more in the spirit of Sec.~\ref{sec:general}). In contrast, evaluating Eqs.~\eqref{eq:speedlimitReplicatorMutation} and~\eqref{eq:upperlowerSL} only requires knowledge of averages and standard deviations.

 The results in this article can be used to discriminate traits with fast or slow evolutionary rates and how different evolutionary forces affect them. Given knowledge of expectation values $\{ \langle f \rangle, \langle A \rangle, \langle A \rangle_\Pi \}$ and standard deviations $\{ \sigma_f, \sigma_A, \sigma^\Pi_A \}$, the right-hand side of Eqs.~(\ref{eq:lowerSL}) and~(\ref{eq:upperSL}) constrain the evolutionary rates of a trait $A$. The rate limits allow mathematically identifying two extremes of evolutionary regimes: (i) The maximum evolution rate of a trait that is robust to mutations, with $\langle A \rangle_\Pi \approx \langle A \rangle$, is bounded by the variabilities in the trait and fitness; (ii) Meanwhile, a population with uniform fitness $f_j \approx f_k$, for which $\sigma_f \approx 0$,  leads to constrained dynamics where $\left| d \langle A \rangle/dt \right| \approx \langle f \rangle \big| \langle A \rangle_\Pi - \langle A \rangle \big|$. Natural selection dominates in case (i), often identified as the weak or neutral selection regime~\cite{WILD2007382},  while mutations dominate in case (ii). I illustrate this in Fig.~\ref{fig:fig2}. (Also see the Supplementary Material for a model illustrating these claims and the inequalities derived in this article.)

We can use Eqs.~(\ref{eq:speedlimitReplicatorMutation}) and~(\ref{eq:upperlowerSL}) to study the evolution rates of quantities often considered to characterize populations. The entropy $S \nobreak \coloneqq \nobreak -\sum_j p_j \ln p_j$,  also known as the Shannon-Wiener index in ecology, has been used to measure the diversity in a population~\cite{diversityentropy}. While $S \nobreak \approx \nobreak 0$ if only one type $k$ occurs with $p_k \nobreak \approx \nobreak 1$, one has $S \nobreak = \nobreak \ln N$ if $N$ types are equally likely to occur. Taking $\{ a_j \nobreak \equiv \nobreak I_j \nobreak \coloneqq  \nobreak -\ln p_j\}$ and using that $\dot S \nobreak = \nobreak -\sum_j \dot p_j \ln p_j$ from conservation of probability, Eq.~(\ref{eq:speedlimitReplicatorMutation}) implies that
\begin{align}
 \left| \frac{dS}{dt} - \langle f \rangle \, S\big(p \|\Pi \big)  \right| &\leq \sigma^\Pi_I \, \sigma_f.
\end{align}
Here, the relative entropy (sometimes called the Kullback–Leibler divergence) $S\big( p \| \Pi \big) \nobreak \coloneqq \nobreak -\sum_j p_j \ln \left( \frac{\Pi_j}{p_j} \right)$ serves as a proxy for the distance between the distribution $p$ and the mutation-driven distribution $\Pi$~\cite{weibull1997evolutionary,karev2010replicator,frank2012InfoTheory,Baez2016}. When $S\big( p \| \Pi \big)$ is small it is hard to distinguish $p$ from $\Pi$. Note that $\sigma^\Pi_I \geq 0$ but it is unbounded from above. In cases with an homogeneous population ($\sigma_f \approx 0$) and finite $\sigma^\Pi_I$, mutations drives variability since the entropy only evolves due to the mismatch between the two distributions, with a rate $\frac{dS}{dt} = \langle f \rangle \, S\big(p \| \Pi \big)$. In the mutation-less case in which $p = \Pi$, Theorem 8 of Ref.~\cite{surprisalvariance} implies that $\sigma^\Pi_I = \sigma_I \leq \sqrt{\tfrac{1}{4} \ln^2(N-1) + 1}$. Then, the maximum entropy rate scales as $\max   \big| \dot S \big|  \nobreak \approx \nobreak  \ln N \, \sigma_f/2$ for large $N$.

As a second example, consider the rate of change of the average fitness function for dynamics that incorporate mutations.
Equation~(\ref{eq:speedlimitReplicatorMutation}) becomes
\begin{align}
\label{eq:speedlimitReplicatorMutationFitness}
\left| \frac{d \langle f \rangle}{dt} - \big\langle  \dot f \big\rangle - \langle f \rangle \Big( \langle f \rangle_\Pi - \langle f \rangle \Big)  \right| \leq \sigma^\Pi_f \, \sigma_f,
\end{align}
which coincides with Eq.~(\ref{eq:thmNatSel}) in the mutation-less regime or for neutral mutations that do not influence the fitness landscape, since then $\langle f \rangle_\Pi = \langle f \rangle$ and $\sigma^\Pi_f = \sigma_f$.
 
Equation~(\ref{eq:speedlimitReplicatorMutationFitness}) imposes the most stringent constraints on evolution when the fitness is completely certain when evaluated in the distribution $p$ or in the mutation-driven distribution $\Pi$, i.e., when $\sigma^\Pi_f = 0$ or $\sigma_f = 0$. This is the neutral selection strong mutation regime (center column of Fig~\ref{fig:fig2}), when types have comparable fitness and natural selection is not a strong driver of evolution. In these cases, dynamics are due to the difference in fitness between the two distributions, and $d\langle f \rangle / dt - \big\langle  \dot f \big\rangle = \langle f \rangle \big( \langle f \rangle_\Pi - \langle f \rangle \big)$, that is, mutations dominate. This also illustrates that beneficial and deleterious mutations can be naturally characterized by $\langle f \rangle_\Pi \nobreak - \nobreak  \langle f \rangle \nobreak \geq \nobreak  0$ and $\langle f \rangle_\Pi \nobreak - \nobreak  \langle f \rangle \nobreak \leq \nobreak  0$, respectively, depending on the change in the average fitness of the population. Note that the contribution to the fitness rate depends on the population's average fitness but not on its standard deviation---Fisher's theorem of natural selection does not hold in this regime. 

 In contrast to the neutral selection regime, faster evolutionary rates are possible for systems with variable fitness landscapes. In this way, rate limits can be used to mathematically formalize evolutionary regimes according to the dynamical forces in action. I will explore this further at the end of the next section.

\begin{figure*}
  \centering   
  \textbf{(a) Strong selection regime \quad \quad\quad \qquad (b) Strong mutation regime \quad \quad\quad \qquad (c) Genetic drift regime} \\
  \vspace{4pt} 
  \includegraphics[trim=00 00 00 00,width=0.32\textwidth]{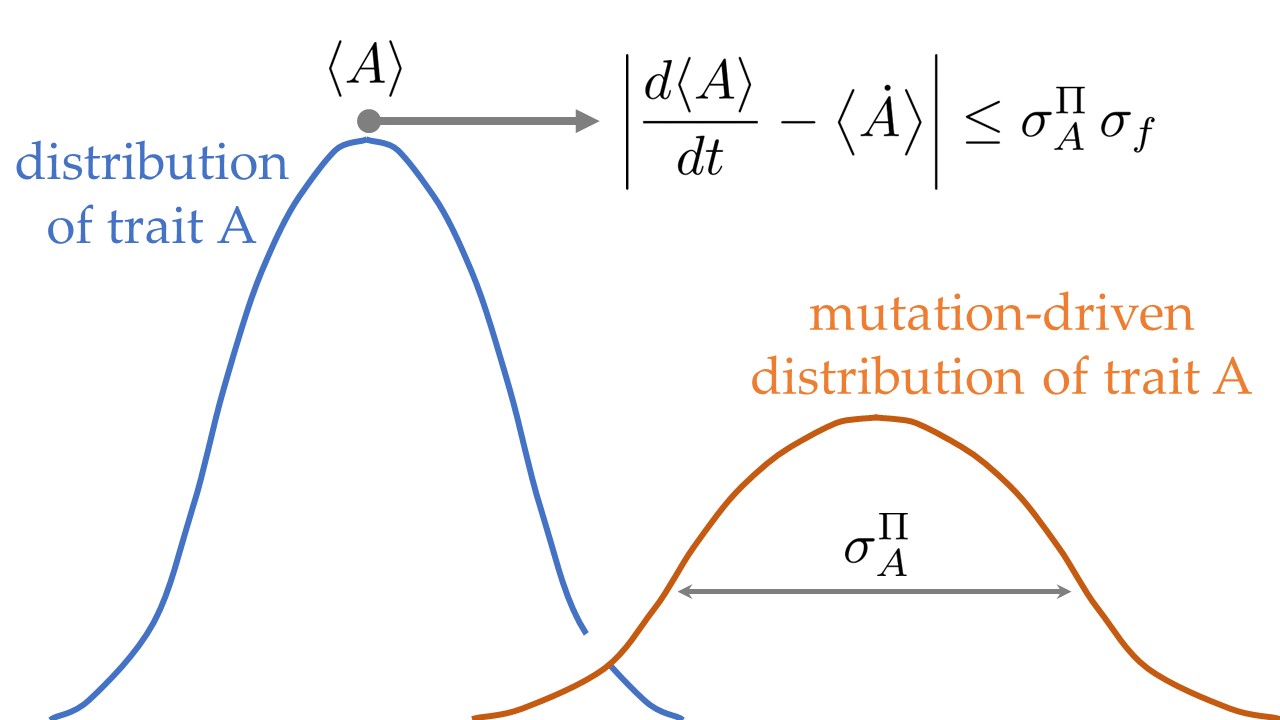}  \, 
  \includegraphics[trim=00 00 00 00,width=0.32\textwidth]{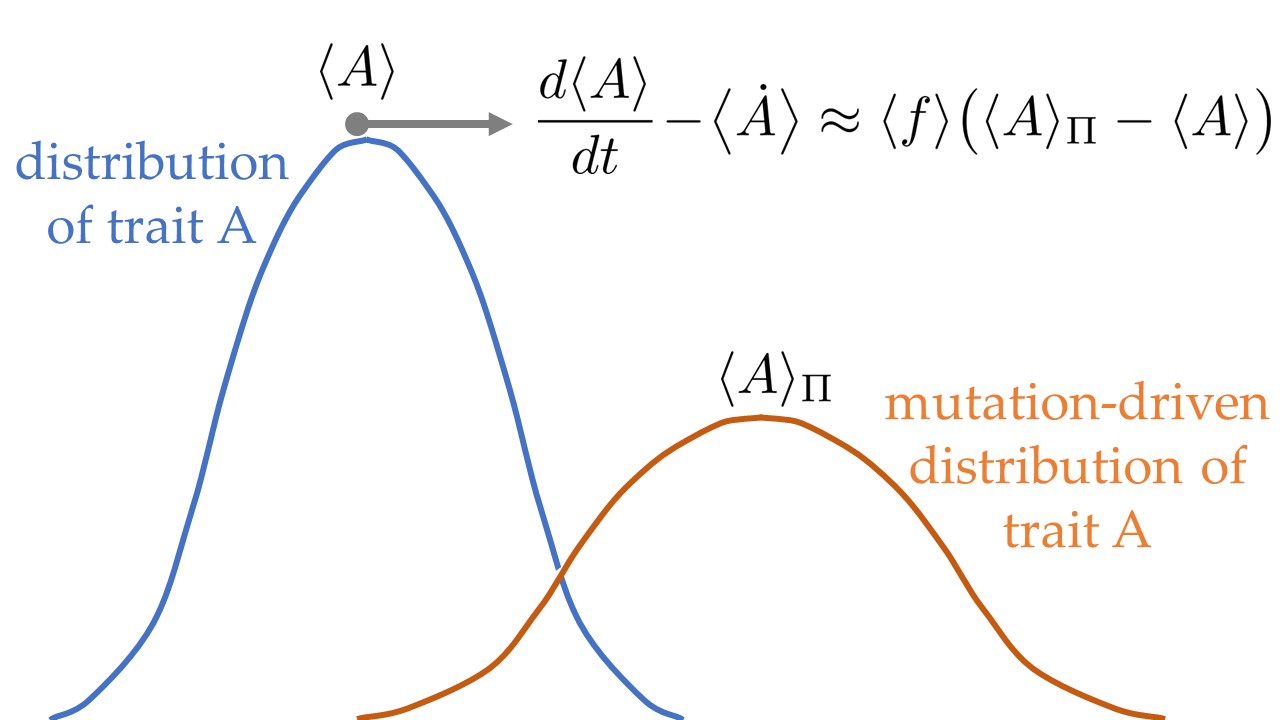}  \, 
  \includegraphics[trim=00 00 00 00,width=0.32\textwidth]{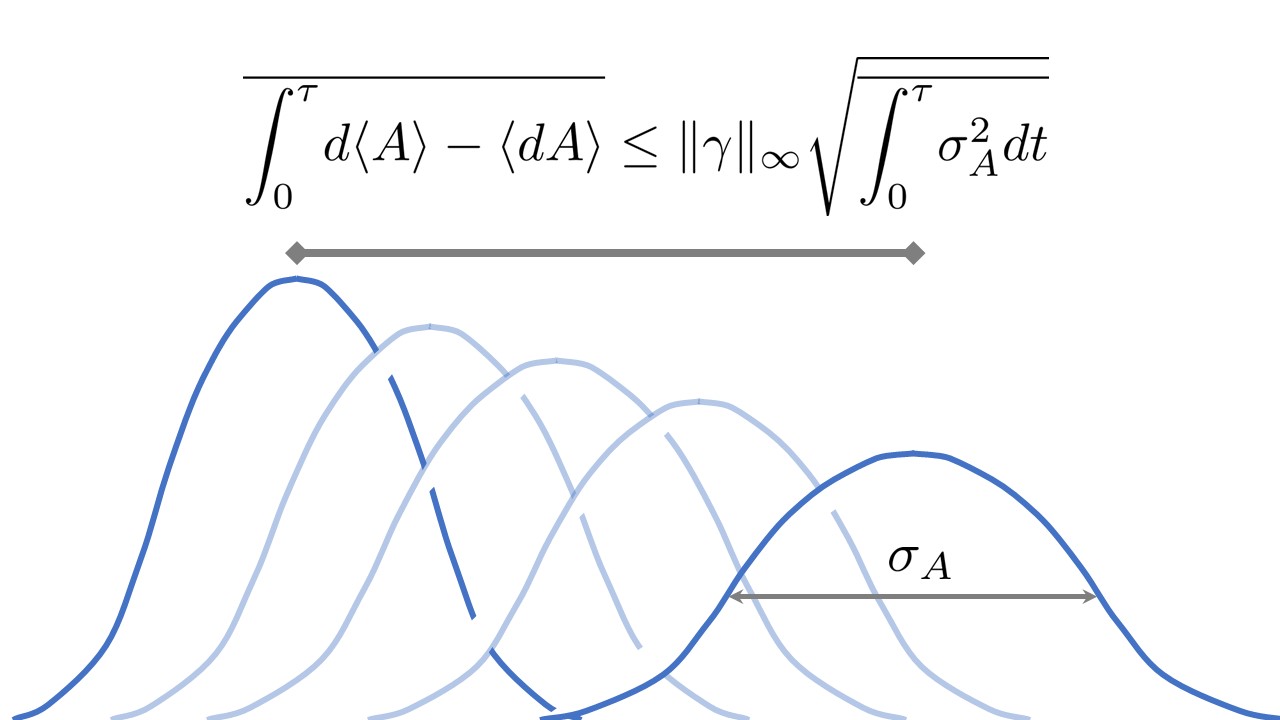} \\
\caption{
\textbf{Discriminating evolutionary regimes.---} 
\label{fig:fig2}
The rate limits derived in this work constrain the evolution of quantifiable traits of populations driven by natural selection, mutations, or genetic drift. The  inequalities depend on (i) the expectation values $\{ \langle f \rangle, \langle A \rangle, \langle A \rangle_\Pi \}$ and (ii) standard deviations $\{ \sigma_f, \sigma_A, \sigma^\Pi_A \}$ of the trait of interest and of the fitness profile of the population, and (iii) on the maximum strength $\|\gamma\|_\infty$ of genetic drift stochastic forces. Here, $\langle A \rangle_\Pi$ and $\sigma^\Pi_A$ are evaluated in the distribution $\Pi_j$ defined in Eq.~(\ref{eq:mutation-drivenProb}), which characterizes type's frequencies if the system only evolved due to mutations. Equation~(\ref{eq:speedlimitStochasticFull}) can be used to mathematically delineate three distinct evolutionary regimes:
\\ \vspace{-8pt} \\
\textcolor{white}{-------}  {\,\,\,[(a) -- Strong selection regime]} When $\sigma^\Pi_A \sigma_f \, \gg \, \big\{ \langle f \rangle  | \langle A \rangle_\Pi \!-\! \langle A \rangle |, \|\gamma\|_\infty \sigma_A \big\}$, the contributions of mutations or genetic drift to the changes in a trait are negligible. This can be identified as the regime where natural selection is the dominant evolutionary force. In it, the maximum rate of a trait is constrained by its standard deviation $\sigma_A^\Pi$ evaluated in the mutation-driven distribution $\Pi$: smaller trait variability implies smaller maximum rates. 
\\ \vspace{-8pt} \\
\textcolor{white}{-------}  {[(b) -- Strong mutation regime]} Natural selection and genetic drift contributions are negligible when $\langle f \rangle | \langle A \rangle_\Pi \nobreak - \nobreak \langle A \rangle  | \, \nobreak \gg \nobreak \, \big\{ \sigma^\Pi_A \sigma_f, \|\gamma\|_\infty \sigma_A \big\}$. Then, the rate of change of a trait is determined by the change induced by the mutation-driven distribution, and Eq.~(\ref{eq:speedlimitReplicatorMutation}) implies that $d\langle A \rangle/dt \nobreak-\nobreak \langle \dot A \rangle  \nobreak\approx\nobreak \langle f \rangle  (\langle A \rangle_\Pi \!-\! \langle A \rangle )$. 
\\ \vspace{-8pt} \\
\textcolor{white}{---------}  {[(c) -- Genetic drift regime]} 
 The variability of a trait determines its susceptibility to changes due to genetic drift. In the regime $\|\gamma\|_\infty \sigma_A \, \gg \, \big\{ \sigma^\Pi_A \sigma_f, \langle f \rangle  | \langle A \rangle_\Pi \!-\! \langle A \rangle  |  \big\}$ where the stochastic force dominates, the total change in a trait over a time $\tau$ is bounded by the integrated variance of the trait and the maximum drift strength.
}
\end{figure*}

 \section{Limits to stochastic evolutionary processes}
 \label{sec:stochastic}
Equations~(\ref{eq:speedlimitReplicatorMutation}) and~(\ref{eq:speedlimitReplicatorMutationFitness}) provide generalizations of Fisher's fundamental theorem of natural selection to replicator-mutator dynamics in terms of universal constraints on evolutionary rates.
However, the replicator and replicator-mutator equations are only simplified deterministic models for the dynamics of infinite populations. In realistic scenarios, stochastic forces typically known as \emph{genetic drift} play a preponderant role in evolutionary processes~\cite{FOSTER1990,
 TrulsenContinuum2021}.

To account for genetic drift, I consider a stochastic replicator-mutator equation,
\begin{align}\label{eq:stochastic}
d p_j =& \sum_k p_k Q_{kj} f_k dt - p_j \langle f \rangle dt  \\
&\qquad\qquad+ p_j \bigg( \gamma_j dW_j - \sum_l \! \gamma_l p_l dW_l \bigg), \nonumber  
\end{align}
as a model for stochastic evolutionary processes. The first two terms coincide with the replicator-mutator Eq.~(\ref{eq:ReplicatorMutation}) and describe natural selection and mutation dynamics. The last term models genetic drift driven by noise terms $dW_j$.

In the stochastic difference equation~(\ref{eq:stochastic}), the population changes by $dp_j$ during a time-step $dt$, over which a stochastic Wiener noise $dW_j$ randomly affects the system. The strength of the genetic drift of population $j$ is characterized by the `volatility parameter' $\gamma_j$~\cite{bedford2009optimization}, while the term $\sum_l\! \gamma_l p_l dW_l$ ensures that the frequencies $p_j$ remain normalized. Following the rules of It\^o calculus, the zero-mean Wiener noises $dW_j$ satisfy $dW_j^2 = dt$ and $\overline{dW_j dW_k} = \delta_{jk} dt$, where $\overline{F}$ represents the average of a function $F(p)$ over realizations of the stochastic noise~\cite{jacobs_stochastic}.

Dynamics are thus driven by natural selection, mutations, and random genetic drift. Note that the latter dominate dynamics for infinitesimally short times~\cite{jacobs_stochastic}. Therefore, in this case, it is more meaningful to focus on integrated changes $\tfrac{1}{\tau}\int_0^\tau \big( d \langle A \rangle \nobreak-\nobreak \langle dA \rangle \big)$ during a time interval $\tau$ rather than on rates of change. The prefactor $\tfrac{1}{\tau}$, which makes the quantity a time-average, ensures the same units as in Eqs.~(\ref{eq:speedlimitReplicator}) and~(\ref{eq:speedlimitReplicatorMutation}).

Then, I prove in the Supplementary Material that, 
\begin{align}
\label{eq:speedlimitStochastic}
&\overline{\left| \frac{1}{\tau} \!\int_0^\tau\! \bigg( d \langle A \rangle - \langle dA \rangle -\sum_{jk} p_k Q_{kj}\big(f_k\!-\!\langle f \rangle \big) a_j dt  \bigg)\right|^2} \nonumber \\
&\qquad \qquad  \qquad \qquad \leq  \frac{\| \gamma \|_\infty^2}{\tau} \, \frac{1}{\tau}\!\int_0^\tau \overline{ \sigma_A^2 } dt, 
\end{align}
where $\| \gamma \|_\infty \coloneqq \max_j {\{\gamma_j\}}$. 
Equation~(\ref{eq:speedlimitStochastic}) shows that the noise-averaged change in $\langle A \rangle$ with respect to the change $\big[ \int_0^\tau \sum_{jk} p_k Q_{kj}\big(f_k\!-\!\langle f \rangle \big) a_j dt \big]$ due to natural selection and mutations, is bounded by the variability $\sigma_A$ and by the maximum strength $\| \gamma \|_\infty$ of the stochastic forces. (One can apply the same techniques I used to derive Eq.~(\ref{eq:speedlimitStochastic}) to other Wiener-noise dynamical models~\cite{FOSTER1990,cabrales2000stochastic,PhysRevE.105.044403}, see the Supplementary Material for one such example.) 

In plain terms, Eq.~(\ref{eq:speedlimitStochastic}) provides a criteria to discriminate situations in which biological quantities are resilient against genetic drift. A trait's high variability can make its dynamics more susceptible to changes due to stochastic forces. In contrast, traits with small variability $\sigma_A \approx 0$ over a period $\tau$ evolve as if natural selection and mutations were the only driving evolutionary forces.

The final result of this paper singles out the contributions of natural selection, mutations, and genetic drift to an inequality on the noise-averaged change of a trait over a time interval $\tau$,
\begin{align}
\label{eq:speedlimitStochasticFull}
    \overline{ \left| \int_0^\tau\! \Big( d\langle A \rangle -  \langle dA \rangle \Big)\right| }  \,\, &\leq \, 
    \overline{\int_0^\tau\! \sigma^\Pi_A \, \sigma_f dt} \\
    &+ \, \overline{ \int_0^\tau \! \langle f \rangle \big| \langle A \rangle_\Pi - \langle A \rangle \big| dt} \nonumber \\
    &+ \, \sqrt{ \overline{ \| \gamma \|_\infty^2  \int_0^\tau \!  \sigma_A^2  dt } }. \nonumber
\end{align}

 There are many ways in which evolutionary regimes can be (often phenomenologically) classified depending on their predominant driving forces on a given model~\cite{WILD2007382,
rouzine_solitary_2003,Desai2007,sniegowski2010beneficial,levy2015quantitative}. 
Equation~(\ref{eq:speedlimitStochasticFull}) yields a formal way to mathematically demarcate such regimes by comparing the relative strengths between the sources that dominate the maximum evolutionary rate:
\vspace{-2pt}
\begin{itemize}
\item[(a)] \emph{strong selection regime}  
\vspace{-4pt}
$$\sigma^\Pi_A \sigma_f \, \gg \, \Big\{ \langle f \rangle \big| \langle A \rangle_\Pi - \langle A \rangle \big| \, , \, \|\gamma\|_\infty \sigma_A \Big\},$$
\vspace{-21pt}
\item[(b)] \emph{strong mutation regime} 
\vspace{-4pt}
$$\langle f \rangle \big| \langle A \rangle_\Pi - \langle A \rangle \big| \, \gg \, \Big\{ \sigma^\Pi_A \sigma_f \, , \, \|\gamma\|_\infty \sigma_A \Big\},$$
\vspace{-21pt}
\item[(c)] \emph{genetic drift regime} 
\vspace{-4pt}
$$  \|\gamma\|_\infty \sigma_A \, \gg \, \Big\{ \sigma^\Pi_A \sigma_f \, , \, \langle f \rangle \big| \langle A \rangle_\Pi - \langle A \rangle \big|  \Big\}. $$
\end{itemize}
\LP{In words, the characteristic regimes depend on the values of three quantities: 
the product of the fitness and trait standard deviations ($\sigma^\Pi_A \sigma_f$), 
the mutation-driven change in a trait ($\langle f \rangle \big| \langle A \rangle_\Pi - \langle A \rangle \big|$), and 
the genetic drift's intensity times the trait standard deviation ($\|\gamma\|_\infty \sigma_A$).
(a)~The strong selection regime occurs when the product of the fitness and trait standard deviations are significantly larger than the other two factors. Under this regime, mutation and genetic drift can be ignored.
(b)~The strong mutation regime occurs when the mutation-driven change in a trait (weighed by the average fitness) is the larger term. In it, natural selection and genetic drift are negligible. 
(c)~Finally, the genetic drift regime is identified by the intensity of the genetic drift  times the trait's standard deviation being the significantly larger term. In it, the trait's evolution is solely due to random events.
}

In each of the three regimes detailed above, dynamics of a trait is provably due to the corresponding biological drive. 
\LP{Note that these are} trait-dependent criteria --- as conveyed by Fig.~\ref{fig:fig2}, certain traits are more responsive than others to a given evolutionary force.

\section{Potential implications to experiments}
\label{sec:experiments}
 In this section, I discuss how the theoretical inequalities derived in Secs.~\ref{sec:general}-\ref{sec:stochastic} could be contrasted with the evolution of concrete biological systems. I hope to bridge the results in this article, which borrow ideas from physics, to questions of interest in biology.

In any experiment where one has access to the frequencies $p_j = n_j/N$ with which types occur and to the variability of a trait, evaluating the results in Sec.~\ref{sec:general} is straightforward. In this case, evaluating the left- and right-hand sides of Eq.~\eqref{eq:speedlimit} from experimental data would inform whether a trait evolves as fast as possible. A trait that saturates the inequality is exploiting all biological resources to evolve as fast as nature allows. 

Evaluating the rate limits on systems driven predominantly by natural selection and mutations (Sec.~\ref{sec:mutations}) will involve purposely designed experiments. 
Some of the quantities in Eq.~\eqref{eq:speedlimitReplicatorMutation} are likely to be directly available from experimental data (e.g., $\langle A \rangle$ and $\langle f \rangle$). However, $\langle A \rangle_\Pi$ and $\sigma_a^\Pi$ are the trait average and standard deviation \emph{evaluated in the mutation-driven frequency distribution $\Pi$} [Eq.~\eqref{eq:mutation-drivenProb}]. Evaluating them would require an experiment where the intensity of mutations can be ``cranked up'' relative to that of natural selection. Experiments with microbial evolution and growth arena (MEGA) plates, introduced in Ref.~\cite{MEGAplate}, could be used for this purpose. Reference~\cite{CARPENTER2023e00227} discusses another method to influence the fitness landscape and favour certain mutations. 
 
 Finally, Sec.~\ref{sec:stochastic} includes the effect of random stochastic influences (genetic drift). Climate change is believed to affect the occurrence and impact of extreme weather events~\cite{franco2020climate,trenberth2015attribution}. Let us consider modeling the effects of extreme events on a population by stochastic forces, as in Sec.~\ref{sec:stochastic}. Then, we can ask: What properties of a biological system make it resilient against the effects of climate change? Relying on the conclusions drawn after Eq.~\eqref{eq:speedlimitStochastic}, one can argue that traits with less variability will remain unaffected by extreme events for longer. This is because the right-hand side of Eq.~\eqref{eq:speedlimitStochastic} is small for traits with low variability. Traits with high variability can be affected more. 
Note that this doesn't necessarily mean that the former case yields more resilience against climate change than the latter: it is possible that populations will need to evolve fast to survive and, if so, Eq.~\eqref{eq:speedlimitStochastic} suggests that variability may help~\cite{poloczanska2016responses,biodivredistribition2017}.

\section{Discussion} 
\label{sec:discussion}
Recent works, mostly within the fields of  quantum physics~\cite{deffner2017quantum,gong2022bounds,LPGP2022,AdC2022Quantum} and classical statistical mechanics~\cite{uffink1999thermodynamic,PhysRevLett.114.158101,PhysRevE.102.062132,horowitz2020thermodynamic,PhysRevX.11.041061,JasonGreenPRR2023}, but also  biology~\cite{ConstraintsSelectionPRR2021,adachi2022universal,ItoUgh2023}, derived  trade-off relations that constrain the dynamics of observables. I find it remarkable that constraints that bound speed, while being extremely general, are saturated in certain paradigmatic cases. In stochastic thermodynamics, for example, the rate at which heat is exchanged with a system and the rate at which a system's entropy changes saturates their speed limits for Gibbs states with (arbitrarily) time-dependent temperature~\cite{Nicholson_2020}. In an unrelated setting, certain quantum algorithms have been shown to compute as fast as allowed by speed limits on quantum annealing~\cite{QSLAnneal}.

Here, replicator dynamics---a model often used to describe mutation-less population dynamics in evolutionary biology---have been found to saturate the rate limit for the fitness of a population (the left and right-hand sides of Eq.~\eqref{eq:speedlimitReplicator} coincide for $A=f$). Higher variability in fitness leads to higher evolutionary rates. This last fact was known by Fisher, who connected the rate of change of the average fitness to the variability in the fitness of a population. 

 More generally, I have shown that connections between the variability in fitness and evolutionary rates hold for dynamics that incorporate mutations and genetic drift. Moreover, these connections can be made for the evolution of arbitrary traits.

Then, the main strengths of our results are that (i) Eq.~(\ref{eq:speedlimitReplicatorMutation}) generalizes the constraints on evolutionary rates to arbitrary quantifiable traits and not just the fitness function, (ii) the inequality in Eq.~(\ref{eq:speedlimitReplicatorMutationFitness}) concisely shows how fitness variability influences dynamics, in the spirit of Fisher's theorem, and that (iii) Eq.~(\ref{eq:speedlimitStochastic}) accounts for stochastic evolutionary forces. See Refs.~\cite{Basener2018,adachi2022universal,ItoUgh2023} for related work generalizing Fisher's theorem of natural selection, and deriving distinct speed limits on evolutionary processes driven by natural selection and mutations. Speed limits that incorporate stochastic dynamics, however, are mostly unexplored.

There is an extensive literature devoted to understanding the factors that limit biological evolutionary rates~\cite{Hoffmann2005limits,RateLimits2011,RateLimits2012,RateLimits2013,hoffmann2014evolutionary,RateLimits2019}. 
 However, the complex nature of this question implies that most of such work is devoted to observations on particular traits within particular types or species, based on experimental observations, and sometimes phenomenological. In contrast, the results in this paper take the role of general mathematical theorems that hold for any biological system whose dynamics can be described by the models considered [Eqs.~(\ref{eq:Replicator}),~(\ref{eq:ReplicatorMutation}), or~(\ref{eq:stochastic})]. The techniques introduced here open a path to derive rate limits for modifications of these stochastic dynamical models, too.

It is worth emphasizing that the main results in this work, which constrain rates for dynamics with mutations [Eqs.~(\ref{eq:speedlimitReplicatorMutation}--\ref{eq:speedlimitReplicatorMutationFitness})] and which discriminate quantities that are resilient against stochastic driving forces [Eqs.~(\ref{eq:speedlimitStochastic}, \ref{eq:speedlimitStochasticFull})], are not implied by the general speed limits derived in Ref.~\cite{Nicholson_2020} nor by recent works generalizing the fundamental theorem of natural selection~\cite{adachi2022universal,ItoUgh2023}. Making use of the structure of the differential equations that model biological evolutionary processes was crucial in deriving the new rate limits. In particular, speed limits on stochastic noisy dynamics such as Eq.~(\ref{eq:stochastic}) are mostly unexplored.
\LP{It would be interesting to study applications of the methods used here to other models of evolutionary processes that account for genetic drift and finite size effects~\cite{CrowGeneticDrift1955,TraulsenReview2009,WrightFisher2016,TrulsenContinuum2021}, other stochastic drives~\cite{TraulsenPNAS2015,VasconcelosPRL2017}, or to quantify the environment's effect on trait plasticity~\cite{berg2010plasticity}}. 
At the same time, while I have argued and illustrated in Fig.~\ref{fig:fig2} how the rate limits are useful in discriminating observable traits in terms of their evolutionary rates, it would be insightful to test these ideas in biological systems from experimental data.

The overarching aim of this work was to mathematically formalize the common knowledge that biological diversity influences evolutionary processes. The results shown here \LP{suggest and} quantify ways in which variability, as measured by the standard deviations of fitness and of the biological traits of interest, \LP{can serve} as a resource by allowing for faster evolution. It is tempting to ponder about the implications to related fields~\cite{whitley1994genetic,EvolutionaryAlgo2010introduction,HODGSON201012}.

\section*{Acknowledgements}
I thank Jos\'e Ignacio Arroyo, Jake Bringewatt, Adolfo del Campo, Lucas Fernandes, Daniel Freire, Jason Green, Schuyler Nicholson, Lodovico Scarpa, and Nicol\'as Rubido for discussions related to this work.
This work was supported by the Beyond Moore’s Law project of the Advanced Simulation and Computing Program at Los Alamos National Laboratory (LANL) managed by Triad National Security, LLC, for the National Nuclear Security Administration of the U.S. Department of Energy under contract 89233218CNA000001.
I also acknowledge support by the DOE Office of Science, Office of Advanced Scientific Computing Research, Accelerated Research for Quantum Computing program, Fundamental Algorithmic Research for Quantum Computing (FAR-QC) project. Work at LANL was carried out under the auspices of the US DOE and NNSA under contract No.~DEAC52-06NA25396.

\bibliography{referencesevol}

\begin{thebibliography}{86}%
\makeatletter
\providecommand \@ifxundefined [1]{%
 \@ifx{#1\undefined}
}%
\providecommand \@ifnum [1]{%
 \ifnum #1\expandafter \@firstoftwo
 \else \expandafter \@secondoftwo
 \fi
}%
\providecommand \@ifx [1]{%
 \ifx #1\expandafter \@firstoftwo
 \else \expandafter \@secondoftwo
 \fi
}%
\providecommand \natexlab [1]{#1}%
\providecommand \enquote  [1]{``#1''}%
\providecommand \bibnamefont  [1]{#1}%
\providecommand \bibfnamefont [1]{#1}%
\providecommand \citenamefont [1]{#1}%
\providecommand \href@noop [0]{\@secondoftwo}%
\providecommand \href [0]{\begingroup \@sanitize@url \@href}%
\providecommand \@href[1]{\@@startlink{#1}\@@href}%
\providecommand \@@href[1]{\endgroup#1\@@endlink}%
\providecommand \@sanitize@url [0]{\catcode `\\12\catcode `\$12\catcode
  `\&12\catcode `\#12\catcode `\^12\catcode `\_12\catcode `\%12\relax}%
\providecommand \@@startlink[1]{}%
\providecommand \@@endlink[0]{}%
\providecommand \url  [0]{\begingroup\@sanitize@url \@url }%
\providecommand \@url [1]{\endgroup\@href {#1}{\urlprefix }}%
\providecommand \urlprefix  [0]{URL }%
\providecommand \Eprint [0]{\href }%
\providecommand \doibase [0]{https://doi.org/}%
\providecommand \selectlanguage [0]{\@gobble}%
\providecommand \bibinfo  [0]{\@secondoftwo}%
\providecommand \bibfield  [0]{\@secondoftwo}%
\providecommand \translation [1]{[#1]}%
\providecommand \BibitemOpen [0]{}%
\providecommand \bibitemStop [0]{}%
\providecommand \bibitemNoStop [0]{.\EOS\space}%
\providecommand \EOS [0]{\spacefactor3000\relax}%
\providecommand \BibitemShut  [1]{\csname bibitem#1\endcsname}%
\let\auto@bib@innerbib\@empty
\bibitem [{\citenamefont {Fisher}(1930)}]{fisher1958genetical}%
  \BibitemOpen
  \bibfield  {author} {\bibinfo {author} {\bibfnamefont {R.~A.}\ \bibnamefont
  {Fisher}},\ }\href {https://doi.org/10.5962/bhl.title.27468} {\emph {\bibinfo
  {title} {The genetical theory of natural selection}}}\ (\bibinfo  {publisher}
  {The Clarendon Press},\ \bibinfo {year} {1930})\BibitemShut {NoStop}%
\bibitem [{\citenamefont {Price}(1972)}]{price1972fisher}%
  \BibitemOpen
  \bibfield  {author} {\bibinfo {author} {\bibfnamefont {G.~R.}\ \bibnamefont
  {Price}},\ }\bibfield  {title} {\bibinfo {title} {Fisher's ‘fundamental
  theorem’ made clear},\ }\href
  {https://doi.org/10.1111/j.1469-1809.1972.tb00764.x} {\bibfield  {journal}
  {\bibinfo  {journal} {Annals of human genetics}\ }\textbf {\bibinfo {volume}
  {36}},\ \bibinfo {pages} {129} (\bibinfo {year} {1972})}\BibitemShut
  {NoStop}%
\bibitem [{\citenamefont {Edwards}(2002)}]{edwards2002fundamental}%
  \BibitemOpen
  \bibfield  {author} {\bibinfo {author} {\bibfnamefont {A.}~\bibnamefont
  {Edwards}},\ }\bibfield  {title} {\bibinfo {title} {The fundamental theorem
  of natural selection},\ }\href {https://doi.org/10.1006/tpbi.2002.1570}
  {\bibfield  {journal} {\bibinfo  {journal} {Theoretical population biology}\
  }\textbf {\bibinfo {volume} {61}},\ \bibinfo {pages} {335} (\bibinfo {year}
  {2002})}\BibitemShut {NoStop}%
\bibitem [{\citenamefont {Baez}(2021)}]{e23111436}%
  \BibitemOpen
  \bibfield  {author} {\bibinfo {author} {\bibfnamefont {J.~C.}\ \bibnamefont
  {Baez}},\ }\bibfield  {title} {\bibinfo {title} {The fundamental theorem of
  natural selection},\ }\href {https://doi.org/10.3390/e23111436} {\bibfield
  {journal} {\bibinfo  {journal} {Entropy}\ }\textbf {\bibinfo {volume} {23}},\
  \bibinfo {pages} {1436} (\bibinfo {year} {2021})}\BibitemShut {NoStop}%
\bibitem [{\citenamefont {Dmitriew}()}]{RateLimits2011}%
  \BibitemOpen
  \bibfield  {author} {\bibinfo {author} {\bibfnamefont {C.~M.}\ \bibnamefont
  {Dmitriew}},\ }\bibfield  {title} {\bibinfo {title} {The evolution of growth
  trajectories: what limits growth rate?},\ }\href
  {https://doi.org/https://doi.org/10.1111/j.1469-185X.2010.00136.x} {\bibfield
   {journal} {\bibinfo  {journal} {Biological Reviews}\ }\textbf {\bibinfo
  {volume} {86}},\ \bibinfo {pages} {97}}\BibitemShut {NoStop}%
\bibitem [{\citenamefont {Blows}\ and\ \citenamefont
  {Hoffmann}(2005)}]{Hoffmann2005limits}%
  \BibitemOpen
  \bibfield  {author} {\bibinfo {author} {\bibfnamefont {M.~W.}\ \bibnamefont
  {Blows}}\ and\ \bibinfo {author} {\bibfnamefont {A.~A.}\ \bibnamefont
  {Hoffmann}},\ }\bibfield  {title} {\bibinfo {title} {A reassessment of
  genetic limits to evolutionary change},\ }\href
  {https://doi.org/10.1890/04-1209} {\bibfield  {journal} {\bibinfo  {journal}
  {Ecology}\ }\textbf {\bibinfo {volume} {86}},\ \bibinfo {pages} {1371}
  (\bibinfo {year} {2005})}\BibitemShut {NoStop}%
\bibitem [{\citenamefont {Hoffmann}\ \emph {et~al.}(2014)\citenamefont
  {Hoffmann}, \citenamefont {Donoghue}, \citenamefont {Levin}, \citenamefont
  {Mackay}, \citenamefont {Rieseberg}, \citenamefont {Travis},\ and\
  \citenamefont {Wray}}]{hoffmann2014evolutionary}%
  \BibitemOpen
  \bibfield  {author} {\bibinfo {author} {\bibfnamefont {A.}~\bibnamefont
  {Hoffmann}}, \bibinfo {author} {\bibfnamefont {M.}~\bibnamefont {Donoghue}},
  \bibinfo {author} {\bibfnamefont {S.}~\bibnamefont {Levin}}, \bibinfo
  {author} {\bibfnamefont {T.}~\bibnamefont {Mackay}}, \bibinfo {author}
  {\bibfnamefont {L.}~\bibnamefont {Rieseberg}}, \bibinfo {author}
  {\bibfnamefont {J.}~\bibnamefont {Travis}},\ and\ \bibinfo {author}
  {\bibfnamefont {G.}~\bibnamefont {Wray}},\ }\bibfield  {title} {\bibinfo
  {title} {Evolutionary limits and constraints},\ }\href
  {https://doi.org/doi:10.1515/9781400848065-034} {\bibfield  {journal}
  {\bibinfo  {journal} {The Princeton guide to evolution}\ ,\ \bibinfo {pages}
  {247}} (\bibinfo {year} {2014})}\BibitemShut {NoStop}%
\bibitem [{\citenamefont {Frean}\ \emph {et~al.}(2013)\citenamefont {Frean},
  \citenamefont {Rainey},\ and\ \citenamefont {Traulsen}}]{RateLimits2013}%
  \BibitemOpen
  \bibfield  {author} {\bibinfo {author} {\bibfnamefont {M.}~\bibnamefont
  {Frean}}, \bibinfo {author} {\bibfnamefont {P.~B.}\ \bibnamefont {Rainey}},\
  and\ \bibinfo {author} {\bibfnamefont {A.}~\bibnamefont {Traulsen}},\
  }\bibfield  {title} {\bibinfo {title} {The effect of population structure on
  the rate of evolution},\ }\href {https://doi.org/10.1098/rspb.2013.0211}
  {\bibfield  {journal} {\bibinfo  {journal} {Proceedings of the Royal Society
  B: Biological Sciences}\ }\textbf {\bibinfo {volume} {280}},\ \bibinfo
  {pages} {20130211} (\bibinfo {year} {2013})}\BibitemShut {NoStop}%
\bibitem [{\citenamefont {Tkadlec}\ \emph {et~al.}(2019)\citenamefont
  {Tkadlec}, \citenamefont {Pavlogiannis}, \citenamefont {Chatterjee},\ and\
  \citenamefont {Nowak}}]{RateLimits2019}%
  \BibitemOpen
  \bibfield  {author} {\bibinfo {author} {\bibfnamefont {J.}~\bibnamefont
  {Tkadlec}}, \bibinfo {author} {\bibfnamefont {A.}~\bibnamefont
  {Pavlogiannis}}, \bibinfo {author} {\bibfnamefont {K.}~\bibnamefont
  {Chatterjee}},\ and\ \bibinfo {author} {\bibfnamefont {M.~A.}\ \bibnamefont
  {Nowak}},\ }\bibfield  {title} {\bibinfo {title} {Population structure
  determines the tradeoff between fixation probability and fixation time},\
  }\href {https://doi.org/10.1038/s42003-019-0373-y} {\bibfield  {journal}
  {\bibinfo  {journal} {Communications biology}\ }\textbf {\bibinfo {volume}
  {2}},\ \bibinfo {pages} {138} (\bibinfo {year} {2019})}\BibitemShut {NoStop}%
\bibitem [{\citenamefont {Evans}\ \emph {et~al.}(2012)\citenamefont {Evans},
  \citenamefont {Jones}, \citenamefont {Boyer}, \citenamefont {Brown},
  \citenamefont {Costa}, \citenamefont {Ernest}, \citenamefont {Fitzgerald},
  \citenamefont {Fortelius}, \citenamefont {Gittleman}, \citenamefont
  {Hamilton} \emph {et~al.}}]{RateLimits2012}%
  \BibitemOpen
  \bibfield  {author} {\bibinfo {author} {\bibfnamefont {A.~R.}\ \bibnamefont
  {Evans}}, \bibinfo {author} {\bibfnamefont {D.}~\bibnamefont {Jones}},
  \bibinfo {author} {\bibfnamefont {A.~G.}\ \bibnamefont {Boyer}}, \bibinfo
  {author} {\bibfnamefont {J.~H.}\ \bibnamefont {Brown}}, \bibinfo {author}
  {\bibfnamefont {D.~P.}\ \bibnamefont {Costa}}, \bibinfo {author}
  {\bibfnamefont {S.~M.}\ \bibnamefont {Ernest}}, \bibinfo {author}
  {\bibfnamefont {E.~M.}\ \bibnamefont {Fitzgerald}}, \bibinfo {author}
  {\bibfnamefont {M.}~\bibnamefont {Fortelius}}, \bibinfo {author}
  {\bibfnamefont {J.~L.}\ \bibnamefont {Gittleman}}, \bibinfo {author}
  {\bibfnamefont {M.~J.}\ \bibnamefont {Hamilton}}, \emph {et~al.},\ }\bibfield
   {title} {\bibinfo {title} {The maximum rate of mammal evolution},\ }\href
  {https://doi.org/10.1073/pnas.1120774109} {\bibfield  {journal} {\bibinfo
  {journal} {Proceedings of the National Academy of Sciences}\ }\textbf
  {\bibinfo {volume} {109}},\ \bibinfo {pages} {4187} (\bibinfo {year}
  {2012})}\BibitemShut {NoStop}%
\bibitem [{\citenamefont {Basener}\ and\ \citenamefont
  {Sanford}(2018)}]{Basener2018}%
  \BibitemOpen
  \bibfield  {author} {\bibinfo {author} {\bibfnamefont {W.~F.}\ \bibnamefont
  {Basener}}\ and\ \bibinfo {author} {\bibfnamefont {J.~C.}\ \bibnamefont
  {Sanford}},\ }\bibfield  {title} {\bibinfo {title} {The fundamental theorem
  of natural selection with mutations},\ }\href
  {https://doi.org/10.1007/s00285-017-1190-x} {\bibfield  {journal} {\bibinfo
  {journal} {Journal of Mathematical Biology}\ }\textbf {\bibinfo {volume}
  {76}},\ \bibinfo {pages} {1589} (\bibinfo {year} {2018})}\BibitemShut
  {NoStop}%
\bibitem [{\citenamefont {Lion}(2018)}]{lion2018theoretical}%
  \BibitemOpen
  \bibfield  {author} {\bibinfo {author} {\bibfnamefont {S.}~\bibnamefont
  {Lion}},\ }\bibfield  {title} {\bibinfo {title} {Theoretical approaches in
  evolutionary ecology: environmental feedback as a unifying perspective},\
  }\href {https://doi.org/10.1086/694865} {\bibfield  {journal} {\bibinfo
  {journal} {The American Naturalist}\ }\textbf {\bibinfo {volume} {191}},\
  \bibinfo {pages} {21} (\bibinfo {year} {2018})}\BibitemShut {NoStop}%
\bibitem [{\citenamefont {Marsland~III}\ \emph {et~al.}(2019)\citenamefont
  {Marsland~III}, \citenamefont {Cui},\ and\ \citenamefont
  {Horowitz}}]{marsland2019thermodynamic}%
  \BibitemOpen
  \bibfield  {author} {\bibinfo {author} {\bibfnamefont {R.}~\bibnamefont
  {Marsland~III}}, \bibinfo {author} {\bibfnamefont {W.}~\bibnamefont {Cui}},\
  and\ \bibinfo {author} {\bibfnamefont {J.~M.}\ \bibnamefont {Horowitz}},\
  }\bibfield  {title} {\bibinfo {title} {The thermodynamic uncertainty relation
  in biochemical oscillations},\ }\href
  {https://doi.org/10.1098/rsif.2019.0098} {\bibfield  {journal} {\bibinfo
  {journal} {Journal of the Royal Society Interface}\ }\textbf {\bibinfo
  {volume} {16}},\ \bibinfo {pages} {20190098} (\bibinfo {year}
  {2019})}\BibitemShut {NoStop}%
\bibitem [{\citenamefont {Genthon}\ and\ \citenamefont
  {Lacoste}(2021)}]{ConstraintsSelectionPRR2021}%
  \BibitemOpen
  \bibfield  {author} {\bibinfo {author} {\bibfnamefont {A.}~\bibnamefont
  {Genthon}}\ and\ \bibinfo {author} {\bibfnamefont {D.}~\bibnamefont
  {Lacoste}},\ }\bibfield  {title} {\bibinfo {title} {Universal constraints on
  selection strength in lineage trees},\ }\href
  {https://doi.org/10.1103/PhysRevResearch.3.023187} {\bibfield  {journal}
  {\bibinfo  {journal} {Phys. Rev. Res.}\ }\textbf {\bibinfo {volume} {3}},\
  \bibinfo {pages} {023187} (\bibinfo {year} {2021})}\BibitemShut {NoStop}%
\bibitem [{\citenamefont {Adachi}\ \emph {et~al.}(2022)\citenamefont {Adachi},
  \citenamefont {Iritani},\ and\ \citenamefont
  {Hamazaki}}]{adachi2022universal}%
  \BibitemOpen
  \bibfield  {author} {\bibinfo {author} {\bibfnamefont {K.}~\bibnamefont
  {Adachi}}, \bibinfo {author} {\bibfnamefont {R.}~\bibnamefont {Iritani}},\
  and\ \bibinfo {author} {\bibfnamefont {R.}~\bibnamefont {Hamazaki}},\
  }\bibfield  {title} {\bibinfo {title} {Universal constraint on nonlinear
  population dynamics},\ }\href {https://doi.org/10.1038/s42005-022-00912-4}
  {\bibfield  {journal} {\bibinfo  {journal} {Communications Physics}\ }\textbf
  {\bibinfo {volume} {5}},\ \bibinfo {pages} {1} (\bibinfo {year}
  {2022})}\BibitemShut {NoStop}%
\bibitem [{\citenamefont {Hoshino}\ \emph {et~al.}(2023)\citenamefont
  {Hoshino}, \citenamefont {Nagayama}, \citenamefont {Yoshimura}, \citenamefont
  {Yamagishi},\ and\ \citenamefont {Ito}}]{ItoUgh2023}%
  \BibitemOpen
  \bibfield  {author} {\bibinfo {author} {\bibfnamefont {M.}~\bibnamefont
  {Hoshino}}, \bibinfo {author} {\bibfnamefont {R.}~\bibnamefont {Nagayama}},
  \bibinfo {author} {\bibfnamefont {K.}~\bibnamefont {Yoshimura}}, \bibinfo
  {author} {\bibfnamefont {J.~F.}\ \bibnamefont {Yamagishi}},\ and\ \bibinfo
  {author} {\bibfnamefont {S.}~\bibnamefont {Ito}},\ }\bibfield  {title}
  {\bibinfo {title} {Geometric speed limit for acceleration by natural
  selection in evolutionary processes},\ }\href
  {https://doi.org/10.1103/PhysRevResearch.5.023127} {\bibfield  {journal}
  {\bibinfo  {journal} {Phys. Rev. Res.}\ }\textbf {\bibinfo {volume} {5}},\
  \bibinfo {pages} {023127} (\bibinfo {year} {2023})}\BibitemShut {NoStop}%
\bibitem [{\citenamefont {Rice}(2004)}]{Rice}%
  \BibitemOpen
  \bibfield  {author} {\bibinfo {author} {\bibfnamefont {S.~H.}\ \bibnamefont
  {Rice}},\ }\href@noop {} {\emph {\bibinfo {title} {Evolutionary Theory:
  Mathematical and conceptual foundations}}},\ \bibinfo {edition} {1st}\ ed.\
  (\bibinfo  {publisher} {Sinauer Associates, Inc.},\ \bibinfo {year}
  {2004})\BibitemShut {NoStop}%
\bibitem [{\citenamefont {Vandermeer}(2010)}]{vandermeer2010populations}%
  \BibitemOpen
  \bibfield  {author} {\bibinfo {author} {\bibfnamefont {J.}~\bibnamefont
  {Vandermeer}},\ }\bibfield  {title} {\bibinfo {title} {How populations grow:
  the exponential and logistic equations},\ }\href@noop {} {\bibfield
  {journal} {\bibinfo  {journal} {Nature Education Knowledge}\ }\textbf
  {\bibinfo {volume} {3}},\ \bibinfo {pages} {15} (\bibinfo {year}
  {2010})}\BibitemShut {NoStop}%
\bibitem [{\citenamefont {Price}(1970)}]{price1970selection}%
  \BibitemOpen
  \bibfield  {author} {\bibinfo {author} {\bibfnamefont {G.~R.}\ \bibnamefont
  {Price}},\ }\bibfield  {title} {\bibinfo {title} {Selection and covariance},\
  }\href {https://doi.org/10.1038/227520a0} {\bibfield  {journal} {\bibinfo
  {journal} {Nature}\ }\textbf {\bibinfo {volume} {227}},\ \bibinfo {pages}
  {520} (\bibinfo {year} {1970})}\BibitemShut {NoStop}%
\bibitem [{\citenamefont {Frank}(2012{\natexlab{a}})}]{frank2012Price}%
  \BibitemOpen
  \bibfield  {author} {\bibinfo {author} {\bibfnamefont {S.~A.}\ \bibnamefont
  {Frank}},\ }\bibfield  {title} {\bibinfo {title} {{Natural selection. IV. The
  price equation}},\ }\href {https://doi.org/10.1111/j.1420-9101.2012.02498.x}
  {\bibfield  {journal} {\bibinfo  {journal} {Journal of evolutionary biology}\
  }\textbf {\bibinfo {volume} {25}},\ \bibinfo {pages} {1002} (\bibinfo {year}
  {2012}{\natexlab{a}})}\BibitemShut {NoStop}%
\bibitem [{\citenamefont {Queller}(2017)}]{queller2017fundamental}%
  \BibitemOpen
  \bibfield  {author} {\bibinfo {author} {\bibfnamefont {D.~C.}\ \bibnamefont
  {Queller}},\ }\bibfield  {title} {\bibinfo {title} {Fundamental theorems of
  evolution},\ }\href {https://doi.org/10.1086/690937} {\bibfield  {journal}
  {\bibinfo  {journal} {The American Naturalist}\ }\textbf {\bibinfo {volume}
  {189}},\ \bibinfo {pages} {345} (\bibinfo {year} {2017})}\BibitemShut
  {NoStop}%
\bibitem [{\citenamefont {Nicholson}\ \emph {et~al.}(2020)\citenamefont
  {Nicholson}, \citenamefont {Garc\'ia-Pintos}, \citenamefont {del Campo},\
  and\ \citenamefont {Green}}]{Nicholson_2020}%
  \BibitemOpen
  \bibfield  {author} {\bibinfo {author} {\bibfnamefont {S.~B.}\ \bibnamefont
  {Nicholson}}, \bibinfo {author} {\bibfnamefont {L.~P.}\ \bibnamefont
  {Garc\'ia-Pintos}}, \bibinfo {author} {\bibfnamefont {A.}~\bibnamefont {del
  Campo}},\ and\ \bibinfo {author} {\bibfnamefont {J.~R.}\ \bibnamefont
  {Green}},\ }\bibfield  {title} {\bibinfo {title} {Time--information
  uncertainty relations in thermodynamics},\ }\href
  {https://doi.org/s41567-020-0981-y} {\bibfield  {journal} {\bibinfo
  {journal} {Nature Physics}\ }\textbf {\bibinfo {volume} {16}},\ \bibinfo
  {pages} {1211} (\bibinfo {year} {2020})}\BibitemShut {NoStop}%
\bibitem [{\citenamefont {Garc\'{\i}a-Pintos}\ \emph
  {et~al.}(2022)\citenamefont {Garc\'{\i}a-Pintos}, \citenamefont {Nicholson},
  \citenamefont {Green}, \citenamefont {del Campo},\ and\ \citenamefont
  {Gorshkov}}]{LPGP2022}%
  \BibitemOpen
  \bibfield  {author} {\bibinfo {author} {\bibfnamefont {L.~P.}\ \bibnamefont
  {Garc\'{\i}a-Pintos}}, \bibinfo {author} {\bibfnamefont {S.~B.}\ \bibnamefont
  {Nicholson}}, \bibinfo {author} {\bibfnamefont {J.~R.}\ \bibnamefont
  {Green}}, \bibinfo {author} {\bibfnamefont {A.}~\bibnamefont {del Campo}},\
  and\ \bibinfo {author} {\bibfnamefont {A.~V.}\ \bibnamefont {Gorshkov}},\
  }\bibfield  {title} {\bibinfo {title} {Unifying quantum and classical speed
  limits on observables},\ }\href {https://doi.org/10.1103/PhysRevX.12.011038}
  {\bibfield  {journal} {\bibinfo  {journal} {Phys. Rev. X}\ }\textbf {\bibinfo
  {volume} {12}},\ \bibinfo {pages} {011038} (\bibinfo {year}
  {2022})}\BibitemShut {NoStop}%
\bibitem [{\citenamefont {Frank}\ and\ \citenamefont
  {Bruggeman}(2020)}]{frank2020fundamental}%
  \BibitemOpen
  \bibfield  {author} {\bibinfo {author} {\bibfnamefont {S.~A.}\ \bibnamefont
  {Frank}}\ and\ \bibinfo {author} {\bibfnamefont {F.~J.}\ \bibnamefont
  {Bruggeman}},\ }\bibfield  {title} {\bibinfo {title} {The fundamental
  equations of change in statistical ensembles and biological populations},\
  }\href {https://doi.org/10.3390/e22121395} {\bibfield  {journal} {\bibinfo
  {journal} {Entropy}\ }\textbf {\bibinfo {volume} {22}},\ \bibinfo {pages}
  {1395} (\bibinfo {year} {2020})}\BibitemShut {NoStop}%
\bibitem [{\citenamefont {Colquhoun}(1971)}]{colquhoun1971lectures}%
  \BibitemOpen
  \bibfield  {author} {\bibinfo {author} {\bibfnamefont {D.}~\bibnamefont
  {Colquhoun}},\ }\href@noop {} {\emph {\bibinfo {title} {Lectures on
  biostatistics: an introduction to statistics with applications in biology and
  medicine}}}\ (\bibinfo  {publisher} {David Colquhoun},\ \bibinfo {year}
  {1971})\BibitemShut {NoStop}%
\bibitem [{\citenamefont {Mandelstam}\ and\ \citenamefont
  {Tamm}(1991)}]{mandelstam1991uncertainty}%
  \BibitemOpen
  \bibfield  {author} {\bibinfo {author} {\bibfnamefont {L.}~\bibnamefont
  {Mandelstam}}\ and\ \bibinfo {author} {\bibfnamefont {I.}~\bibnamefont
  {Tamm}},\ }\bibfield  {title} {\bibinfo {title} {The uncertainty relation
  between energy and time in non-relativistic quantum mechanics},\ }in\ \href
  {https://doi.org/10.1007/978-3-642-74626-0_8} {\emph {\bibinfo {booktitle}
  {Selected papers}}}\ (\bibinfo  {publisher} {Springer},\ \bibinfo {year}
  {1991})\ pp.\ \bibinfo {pages} {115--123}\BibitemShut {NoStop}%
\bibitem [{\citenamefont {Deffner}\ and\ \citenamefont
  {Campbell}(2017)}]{deffner2017quantum}%
  \BibitemOpen
  \bibfield  {author} {\bibinfo {author} {\bibfnamefont {S.}~\bibnamefont
  {Deffner}}\ and\ \bibinfo {author} {\bibfnamefont {S.}~\bibnamefont
  {Campbell}},\ }\bibfield  {title} {\bibinfo {title} {Quantum speed limits:
  from {Heisenberg’s} uncertainty principle to optimal quantum control},\
  }\href {https://doi.org/10.1088/1751-8121/aa86c6} {\bibfield  {journal}
  {\bibinfo  {journal} {Journal of Physics A: Mathematical and Theoretical}\
  }\textbf {\bibinfo {volume} {50}},\ \bibinfo {pages} {453001} (\bibinfo
  {year} {2017})}\BibitemShut {NoStop}%
\bibitem [{\citenamefont {Cram{\'e}r}(1999)}]{cramer1999mathematical}%
  \BibitemOpen
  \bibfield  {author} {\bibinfo {author} {\bibfnamefont {H.}~\bibnamefont
  {Cram{\'e}r}},\ }\href@noop {} {\emph {\bibinfo {title} {Mathematical methods
  of statistics}}},\ Vol.~\bibinfo {volume} {43}\ (\bibinfo  {publisher}
  {Princeton university press},\ \bibinfo {year} {1999})\BibitemShut {NoStop}%
\bibitem [{\citenamefont {{Maynard Smith}}(1974)}]{MAYNARDSMITH1974209}%
  \BibitemOpen
  \bibfield  {author} {\bibinfo {author} {\bibfnamefont {J.}~\bibnamefont
  {{Maynard Smith}}},\ }\bibfield  {title} {\bibinfo {title} {The theory of
  games and the evolution of animal conflicts},\ }\href
  {https://doi.org/10.1016/0022-5193(74)90110-6} {\bibfield  {journal}
  {\bibinfo  {journal} {Journal of Theoretical Biology}\ }\textbf {\bibinfo
  {volume} {47}},\ \bibinfo {pages} {209} (\bibinfo {year} {1974})}\BibitemShut
  {NoStop}%
\bibitem [{\citenamefont {Schuster}\ and\ \citenamefont
  {Sigmund}(1983)}]{SCHUSTER1983533}%
  \BibitemOpen
  \bibfield  {author} {\bibinfo {author} {\bibfnamefont {P.}~\bibnamefont
  {Schuster}}\ and\ \bibinfo {author} {\bibfnamefont {K.}~\bibnamefont
  {Sigmund}},\ }\bibfield  {title} {\bibinfo {title} {Replicator dynamics},\
  }\href {https://doi.org/10.1016/0022-5193(83)90445-9} {\bibfield  {journal}
  {\bibinfo  {journal} {Journal of Theoretical Biology}\ }\textbf {\bibinfo
  {volume} {100}},\ \bibinfo {pages} {533} (\bibinfo {year}
  {1983})}\BibitemShut {NoStop}%
\bibitem [{\citenamefont {Sato}\ and\ \citenamefont
  {Crutchfield}(2003)}]{CrutchfieldPRE2003}%
  \BibitemOpen
  \bibfield  {author} {\bibinfo {author} {\bibfnamefont {Y.}~\bibnamefont
  {Sato}}\ and\ \bibinfo {author} {\bibfnamefont {J.~P.}\ \bibnamefont
  {Crutchfield}},\ }\bibfield  {title} {\bibinfo {title} {Coupled replicator
  equations for the dynamics of learning in multiagent systems},\ }\href
  {https://doi.org/10.1103/PhysRevE.67.015206} {\bibfield  {journal} {\bibinfo
  {journal} {Phys. Rev. E}\ }\textbf {\bibinfo {volume} {67}},\ \bibinfo
  {pages} {015206} (\bibinfo {year} {2003})}\BibitemShut {NoStop}%
\bibitem [{\citenamefont {Safarzy{\'{n}}ska}\ and\ \citenamefont {van~den
  Bergh}(2010)}]{ReplicatorEcon2010}%
  \BibitemOpen
  \bibfield  {author} {\bibinfo {author} {\bibfnamefont {K.}~\bibnamefont
  {Safarzy{\'{n}}ska}}\ and\ \bibinfo {author} {\bibfnamefont {J.~C. J.~M.}\
  \bibnamefont {van~den Bergh}},\ }\bibfield  {title} {\bibinfo {title}
  {Evolutionary models in economics: a survey of methods and building blocks},\
  }\href {https://doi.org/10.1007/s00191-009-0153-9} {\bibfield  {journal}
  {\bibinfo  {journal} {Journal of Evolutionary Economics}\ }\textbf {\bibinfo
  {volume} {20}},\ \bibinfo {pages} {329} (\bibinfo {year} {2010})}\BibitemShut
  {NoStop}%
\bibitem [{\citenamefont {Cressman}\ and\ \citenamefont
  {Tao}(2014)}]{ReplicatorPNAS2014}%
  \BibitemOpen
  \bibfield  {author} {\bibinfo {author} {\bibfnamefont {R.}~\bibnamefont
  {Cressman}}\ and\ \bibinfo {author} {\bibfnamefont {Y.}~\bibnamefont {Tao}},\
  }\bibfield  {title} {\bibinfo {title} {The replicator equation and other game
  dynamics},\ }\href {https://doi.org/10.1073/pnas.1400823111} {\bibfield
  {journal} {\bibinfo  {journal} {Proceedings of the National Academy of
  Sciences}\ }\textbf {\bibinfo {volume} {111}},\ \bibinfo {pages} {10810}
  (\bibinfo {year} {2014})}\BibitemShut {NoStop}%
\bibitem [{\citenamefont {Bloembergen}\ \emph {et~al.}(2015)\citenamefont
  {Bloembergen}, \citenamefont {Tuyls}, \citenamefont {Hennes},\ and\
  \citenamefont {Kaisers}}]{bloembergen2015evolutionary}%
  \BibitemOpen
  \bibfield  {author} {\bibinfo {author} {\bibfnamefont {D.}~\bibnamefont
  {Bloembergen}}, \bibinfo {author} {\bibfnamefont {K.}~\bibnamefont {Tuyls}},
  \bibinfo {author} {\bibfnamefont {D.}~\bibnamefont {Hennes}},\ and\ \bibinfo
  {author} {\bibfnamefont {M.}~\bibnamefont {Kaisers}},\ }\bibfield  {title}
  {\bibinfo {title} {Evolutionary dynamics of multi-agent learning: A survey},\
  }\href {https://doi.org/10.1613/jair.4818} {\bibfield  {journal} {\bibinfo
  {journal} {Journal of Artificial Intelligence Research}\ }\textbf {\bibinfo
  {volume} {53}},\ \bibinfo {pages} {659} (\bibinfo {year} {2015})}\BibitemShut
  {NoStop}%
\bibitem [{\citenamefont {Smith}(1998)}]{smith1998evolutionary}%
  \BibitemOpen
  \bibfield  {author} {\bibinfo {author} {\bibfnamefont {J.~M.}\ \bibnamefont
  {Smith}},\ }\href@noop {} {\emph {\bibinfo {title} {Evolutionary genetics}}}\
  (\bibinfo  {publisher} {Oxford University Press},\ \bibinfo {year}
  {1998})\BibitemShut {NoStop}%
\bibitem [{\citenamefont {Naeem}\ \emph {et~al.}(2012)\citenamefont {Naeem},
  \citenamefont {Duffy},\ and\ \citenamefont {Zavaleta}}]{naeem2012functions}%
  \BibitemOpen
  \bibfield  {author} {\bibinfo {author} {\bibfnamefont {S.}~\bibnamefont
  {Naeem}}, \bibinfo {author} {\bibfnamefont {J.~E.}\ \bibnamefont {Duffy}},\
  and\ \bibinfo {author} {\bibfnamefont {E.}~\bibnamefont {Zavaleta}},\
  }\bibfield  {title} {\bibinfo {title} {The functions of biological diversity
  in an age of extinction},\ }\href {https://doi.org/10.1126/science.1215855}
  {\bibfield  {journal} {\bibinfo  {journal} {Science}\ }\textbf {\bibinfo
  {volume} {336}},\ \bibinfo {pages} {1401} (\bibinfo {year}
  {2012})}\BibitemShut {NoStop}%
\bibitem [{\citenamefont {Heywood}(2005)}]{heywood2005exact}%
  \BibitemOpen
  \bibfield  {author} {\bibinfo {author} {\bibfnamefont {J.~S.}\ \bibnamefont
  {Heywood}},\ }\bibfield  {title} {\bibinfo {title} {An exact form of the
  breeder's equation for the evolution of a quantitative trait under natural
  selection},\ }\href {https://doi.org/10.1111/j.0014-3820.2005.tb00939.x}
  {\bibfield  {journal} {\bibinfo  {journal} {Evolution}\ }\textbf {\bibinfo
  {volume} {59}},\ \bibinfo {pages} {2287} (\bibinfo {year}
  {2005})}\BibitemShut {NoStop}%
\bibitem [{\citenamefont {Morrissey}\ \emph {et~al.}(2010)\citenamefont
  {Morrissey}, \citenamefont {Kruuk},\ and\ \citenamefont
  {Wilson}}]{morrissey2010danger}%
  \BibitemOpen
  \bibfield  {author} {\bibinfo {author} {\bibfnamefont {M.~B.}\ \bibnamefont
  {Morrissey}}, \bibinfo {author} {\bibfnamefont {L.~E.}\ \bibnamefont
  {Kruuk}},\ and\ \bibinfo {author} {\bibfnamefont {A.~J.}\ \bibnamefont
  {Wilson}},\ }\bibfield  {title} {\bibinfo {title} {The danger of applying the
  breeder's equation in observational studies of natural populations},\ }\href
  {https://doi.org/10.1111/j.1420-9101.2010.02084.x} {\bibfield  {journal}
  {\bibinfo  {journal} {Journal of evolutionary biology}\ }\textbf {\bibinfo
  {volume} {23}},\ \bibinfo {pages} {2277} (\bibinfo {year}
  {2010})}\BibitemShut {NoStop}%
\bibitem [{\citenamefont {Gregory}(2009)}]{gregory2009understanding}%
  \BibitemOpen
  \bibfield  {author} {\bibinfo {author} {\bibfnamefont {T.~R.}\ \bibnamefont
  {Gregory}},\ }\bibfield  {title} {\bibinfo {title} {Understanding natural
  selection: essential concepts and common misconceptions},\ }\href
  {https://doi.org/10.1007/s12052-009-0128-1} {\bibfield  {journal} {\bibinfo
  {journal} {Evolution: Education and outreach}\ }\textbf {\bibinfo {volume}
  {2}},\ \bibinfo {pages} {156} (\bibinfo {year} {2009})}\BibitemShut {NoStop}%
\bibitem [{\citenamefont {Carlin}(2011)}]{carlin2011mutations}%
  \BibitemOpen
  \bibfield  {author} {\bibinfo {author} {\bibfnamefont {J.}~\bibnamefont
  {Carlin}},\ }\bibfield  {title} {\bibinfo {title} {Mutations are the raw
  materials of evolution},\ }\href@noop {} {\bibfield  {journal} {\bibinfo
  {journal} {Nature Education Knowledge}\ }\textbf {\bibinfo {volume} {3}},\
  \bibinfo {pages} {10} (\bibinfo {year} {2011})}\BibitemShut {NoStop}%
\bibitem [{\citenamefont {Hershberg}(2015)}]{hershberg2015mutation}%
  \BibitemOpen
  \bibfield  {author} {\bibinfo {author} {\bibfnamefont {R.}~\bibnamefont
  {Hershberg}},\ }\bibfield  {title} {\bibinfo {title} {Mutation—the engine
  of evolution: studying mutation and its role in the evolution of bacteria},\
  }\href {https://doi.org/10.1101/cshperspect.a018077} {\bibfield  {journal}
  {\bibinfo  {journal} {Cold Spring Harbor perspectives in biology}\ }\textbf
  {\bibinfo {volume} {7}},\ \bibinfo {pages} {a018077} (\bibinfo {year}
  {2015})}\BibitemShut {NoStop}%
\bibitem [{\citenamefont {Eigen}\ and\ \citenamefont
  {Schuster}(1982)}]{eigen1982stages}%
  \BibitemOpen
  \bibfield  {author} {\bibinfo {author} {\bibfnamefont {M.}~\bibnamefont
  {Eigen}}\ and\ \bibinfo {author} {\bibfnamefont {P.}~\bibnamefont
  {Schuster}},\ }\bibfield  {title} {\bibinfo {title} {Stages of emerging
  life—five principles of early organization},\ }\href
  {https://doi.org/10.1007/BF02100223} {\bibfield  {journal} {\bibinfo
  {journal} {Journal of molecular evolution}\ }\textbf {\bibinfo {volume}
  {19}},\ \bibinfo {pages} {47} (\bibinfo {year} {1982})}\BibitemShut {NoStop}%
\bibitem [{\citenamefont {Hofbauer}(1985)}]{Hofbauer1985}%
  \BibitemOpen
  \bibfield  {author} {\bibinfo {author} {\bibfnamefont {J.}~\bibnamefont
  {Hofbauer}},\ }\bibfield  {title} {\bibinfo {title} {The selection mutation
  equation},\ }\href {https://doi.org/10.1007/BF00276557} {\bibfield  {journal}
  {\bibinfo  {journal} {Journal of Mathematical Biology}\ }\textbf {\bibinfo
  {volume} {23}},\ \bibinfo {pages} {41} (\bibinfo {year} {1985})}\BibitemShut
  {NoStop}%
\bibitem [{\citenamefont {Wilke}(2005)}]{wilke2005quasispecies}%
  \BibitemOpen
  \bibfield  {author} {\bibinfo {author} {\bibfnamefont {C.~O.}\ \bibnamefont
  {Wilke}},\ }\bibfield  {title} {\bibinfo {title} {Quasispecies theory in the
  context of population genetics},\ }\href
  {https://doi.org/10.1186/1471-2148-5-44} {\bibfield  {journal} {\bibinfo
  {journal} {BMC evolutionary biology}\ }\textbf {\bibinfo {volume} {5}},\
  \bibinfo {pages} {1} (\bibinfo {year} {2005})}\BibitemShut {NoStop}%
\bibitem [{\citenamefont {Nowak}(2006)}]{nowak2006evolutionary}%
  \BibitemOpen
  \bibfield  {author} {\bibinfo {author} {\bibfnamefont {M.~A.}\ \bibnamefont
  {Nowak}},\ }\href {https://doi.org/10.2307/j.ctvjghw98} {\emph {\bibinfo
  {title} {Evolutionary dynamics: exploring the equations of life}}}\ (\bibinfo
   {publisher} {Harvard university press},\ \bibinfo {year} {2006})\BibitemShut
  {NoStop}%
\bibitem [{\citenamefont {Wild}\ and\ \citenamefont
  {Traulsen}(2007)}]{WILD2007382}%
  \BibitemOpen
  \bibfield  {author} {\bibinfo {author} {\bibfnamefont {G.}~\bibnamefont
  {Wild}}\ and\ \bibinfo {author} {\bibfnamefont {A.}~\bibnamefont
  {Traulsen}},\ }\bibfield  {title} {\bibinfo {title} {The different limits of
  weak selection and the evolutionary dynamics of finite populations},\ }\href
  {https://doi.org/https://doi.org/10.1016/j.jtbi.2007.03.015} {\bibfield
  {journal} {\bibinfo  {journal} {Journal of Theoretical Biology}\ }\textbf
  {\bibinfo {volume} {247}},\ \bibinfo {pages} {382} (\bibinfo {year}
  {2007})}\BibitemShut {NoStop}%
\bibitem [{\citenamefont {Spellerberg}\ and\ \citenamefont
  {Fedor}(2003)}]{diversityentropy}%
  \BibitemOpen
  \bibfield  {author} {\bibinfo {author} {\bibfnamefont {I.~F.}\ \bibnamefont
  {Spellerberg}}\ and\ \bibinfo {author} {\bibfnamefont {P.~J.}\ \bibnamefont
  {Fedor}},\ }\bibfield  {title} {\bibinfo {title} {A tribute to {Claude}
  {Shannon} (1916--2001) and a plea for more rigorous use of species richness,
  species diversity and the ‘{Shannon}--{Wiener}’ index},\ }\href@noop {}
  {\bibfield  {journal} {\bibinfo  {journal} {Global ecology and biogeography}\
  }\textbf {\bibinfo {volume} {12}},\ \bibinfo {pages} {177} (\bibinfo {year}
  {2003})}\BibitemShut {NoStop}%
\bibitem [{\citenamefont {Weibull}(1997)}]{weibull1997evolutionary}%
  \BibitemOpen
  \bibfield  {author} {\bibinfo {author} {\bibfnamefont {J.~W.}\ \bibnamefont
  {Weibull}},\ }\href@noop {} {\emph {\bibinfo {title} {Evolutionary game
  theory}}}\ (\bibinfo  {publisher} {MIT press},\ \bibinfo {year}
  {1997})\BibitemShut {NoStop}%
\bibitem [{\citenamefont {Karev}(2010)}]{karev2010replicator}%
  \BibitemOpen
  \bibfield  {author} {\bibinfo {author} {\bibfnamefont {G.~P.}\ \bibnamefont
  {Karev}},\ }\bibfield  {title} {\bibinfo {title} {Replicator equations and
  the principle of minimal production of information},\ }\href
  {https://doi.org/10.1007/s11538-009-9484-9} {\bibfield  {journal} {\bibinfo
  {journal} {Bulletin of mathematical biology}\ }\textbf {\bibinfo {volume}
  {72}},\ \bibinfo {pages} {1124} (\bibinfo {year} {2010})}\BibitemShut
  {NoStop}%
\bibitem [{\citenamefont {Frank}(2012{\natexlab{b}})}]{frank2012InfoTheory}%
  \BibitemOpen
  \bibfield  {author} {\bibinfo {author} {\bibfnamefont {S.~A.}\ \bibnamefont
  {Frank}},\ }\bibfield  {title} {\bibinfo {title} {{Natural selection. V. How
  to read the fundamental equations of evolutionary change in terms of
  information theory}},\ }\href {https://doi.org/10.1111/jeb.12010} {\bibfield
  {journal} {\bibinfo  {journal} {Journal of evolutionary biology}\ }\textbf
  {\bibinfo {volume} {25}},\ \bibinfo {pages} {2377} (\bibinfo {year}
  {2012}{\natexlab{b}})}\BibitemShut {NoStop}%
\bibitem [{\citenamefont {Baez}\ and\ \citenamefont
  {Pollard}(2016)}]{Baez2016}%
  \BibitemOpen
  \bibfield  {author} {\bibinfo {author} {\bibfnamefont {J.~C.}\ \bibnamefont
  {Baez}}\ and\ \bibinfo {author} {\bibfnamefont {B.~S.}\ \bibnamefont
  {Pollard}},\ }\bibfield  {title} {\bibinfo {title} {Relative entropy in
  biological systems},\ }\href {https://doi.org/10.3390/e18020046} {\bibfield
  {journal} {\bibinfo  {journal} {Entropy}\ }\textbf {\bibinfo {volume} {18}},\
  \bibinfo {pages} {46} (\bibinfo {year} {2016})}\BibitemShut {NoStop}%
\bibitem [{\citenamefont {Reeb}\ and\ \citenamefont
  {Wolf}(2015)}]{surprisalvariance}%
  \BibitemOpen
  \bibfield  {author} {\bibinfo {author} {\bibfnamefont {D.}~\bibnamefont
  {Reeb}}\ and\ \bibinfo {author} {\bibfnamefont {M.~M.}\ \bibnamefont
  {Wolf}},\ }\bibfield  {title} {\bibinfo {title} {Tight bound on relative
  entropy by entropy difference},\ }\href
  {https://doi.org/10.1109/TIT.2014.2387822} {\bibfield  {journal} {\bibinfo
  {journal} {IEEE Transactions on Information Theory}\ }\textbf {\bibinfo
  {volume} {61}},\ \bibinfo {pages} {1458} (\bibinfo {year}
  {2015})}\BibitemShut {NoStop}%
\bibitem [{\citenamefont {Foster}\ and\ \citenamefont
  {Young}(1990)}]{FOSTER1990}%
  \BibitemOpen
  \bibfield  {author} {\bibinfo {author} {\bibfnamefont {D.}~\bibnamefont
  {Foster}}\ and\ \bibinfo {author} {\bibfnamefont {P.}~\bibnamefont {Young}},\
  }\bibfield  {title} {\bibinfo {title} {Stochastic evolutionary game
  dynamics},\ }\href
  {https://doi.org/https://doi.org/10.1016/0040-5809(90)90011-J} {\bibfield
  {journal} {\bibinfo  {journal} {Theoretical Population Biology}\ }\textbf
  {\bibinfo {volume} {38}},\ \bibinfo {pages} {219} (\bibinfo {year}
  {1990})}\BibitemShut {NoStop}%
\bibitem [{\citenamefont {Czuppon}\ and\ \citenamefont
  {Traulsen}(2021)}]{TrulsenContinuum2021}%
  \BibitemOpen
  \bibfield  {author} {\bibinfo {author} {\bibfnamefont {P.}~\bibnamefont
  {Czuppon}}\ and\ \bibinfo {author} {\bibfnamefont {A.}~\bibnamefont
  {Traulsen}},\ }\bibfield  {title} {\bibinfo {title} {Understanding
  evolutionary and ecological dynamics using a continuum limit},\ }\href
  {https://doi.org/https://doi.org/10.1002/ece3.7205} {\bibfield  {journal}
  {\bibinfo  {journal} {Ecology and Evolution}\ }\textbf {\bibinfo {volume}
  {11}},\ \bibinfo {pages} {5857} (\bibinfo {year} {2021})}\BibitemShut
  {NoStop}%
\bibitem [{\citenamefont {Bedford}\ and\ \citenamefont
  {Hartl}(2009)}]{bedford2009optimization}%
  \BibitemOpen
  \bibfield  {author} {\bibinfo {author} {\bibfnamefont {T.}~\bibnamefont
  {Bedford}}\ and\ \bibinfo {author} {\bibfnamefont {D.~L.}\ \bibnamefont
  {Hartl}},\ }\bibfield  {title} {\bibinfo {title} {Optimization of gene
  expression by natural selection},\ }\href
  {https://doi.org/10.1073/pnas.0812009106} {\bibfield  {journal} {\bibinfo
  {journal} {Proceedings of the National Academy of Sciences}\ }\textbf
  {\bibinfo {volume} {106}},\ \bibinfo {pages} {1133} (\bibinfo {year}
  {2009})}\BibitemShut {NoStop}%
\bibitem [{\citenamefont {Jacobs}(2010)}]{jacobs_stochastic}%
  \BibitemOpen
  \bibfield  {author} {\bibinfo {author} {\bibfnamefont {K.}~\bibnamefont
  {Jacobs}},\ }\href {https://doi.org/10.1017/CBO9780511815980} {\emph
  {\bibinfo {title} {Stochastic Processes for Physicists: Understanding Noisy
  Systems}}}\ (\bibinfo  {publisher} {Cambridge University Press},\ \bibinfo
  {year} {2010})\BibitemShut {NoStop}%
\bibitem [{\citenamefont {Cabrales}(2000)}]{cabrales2000stochastic}%
  \BibitemOpen
  \bibfield  {author} {\bibinfo {author} {\bibfnamefont {A.}~\bibnamefont
  {Cabrales}},\ }\bibfield  {title} {\bibinfo {title} {Stochastic replicator
  dynamics},\ }\href {https://doi.org/10.1111/1468-2354.00071} {\bibfield
  {journal} {\bibinfo  {journal} {International Economic Review}\ }\textbf
  {\bibinfo {volume} {41}},\ \bibinfo {pages} {451} (\bibinfo {year}
  {2000})}\BibitemShut {NoStop}%
\bibitem [{\citenamefont {Feng}\ \emph {et~al.}(2022)\citenamefont {Feng},
  \citenamefont {Li}, \citenamefont {Zheng}, \citenamefont {Lessard},\ and\
  \citenamefont {Tao}}]{PhysRevE.105.044403}%
  \BibitemOpen
  \bibfield  {author} {\bibinfo {author} {\bibfnamefont {T.-J.}\ \bibnamefont
  {Feng}}, \bibinfo {author} {\bibfnamefont {C.}~\bibnamefont {Li}}, \bibinfo
  {author} {\bibfnamefont {X.-D.}\ \bibnamefont {Zheng}}, \bibinfo {author}
  {\bibfnamefont {S.}~\bibnamefont {Lessard}},\ and\ \bibinfo {author}
  {\bibfnamefont {Y.}~\bibnamefont {Tao}},\ }\bibfield  {title} {\bibinfo
  {title} {Stochastic replicator dynamics and evolutionary stability},\ }\href
  {https://doi.org/10.1103/PhysRevE.105.044403} {\bibfield  {journal} {\bibinfo
   {journal} {Phys. Rev. E}\ }\textbf {\bibinfo {volume} {105}},\ \bibinfo
  {pages} {044403} (\bibinfo {year} {2022})}\BibitemShut {NoStop}%
\bibitem [{\citenamefont {Rouzine}\ \emph {et~al.}(2003)\citenamefont
  {Rouzine}, \citenamefont {Wakeley},\ and\ \citenamefont
  {Coffin}}]{rouzine_solitary_2003}%
  \BibitemOpen
  \bibfield  {author} {\bibinfo {author} {\bibfnamefont {I.~M.}\ \bibnamefont
  {Rouzine}}, \bibinfo {author} {\bibfnamefont {J.}~\bibnamefont {Wakeley}},\
  and\ \bibinfo {author} {\bibfnamefont {J.~M.}\ \bibnamefont {Coffin}},\
  }\bibfield  {title} {\bibinfo {title} {The solitary wave of asexual
  evolution},\ }\href {https://doi.org/10.1073/pnas.242719299} {\bibfield
  {journal} {\bibinfo  {journal} {Proceedings of the National Academy of
  Sciences}\ }\textbf {\bibinfo {volume} {100}},\ \bibinfo {pages} {587}
  (\bibinfo {year} {2003})}\BibitemShut {NoStop}%
\bibitem [{\citenamefont {Desai}\ and\ \citenamefont
  {Fisher}(2007)}]{Desai2007}%
  \BibitemOpen
  \bibfield  {author} {\bibinfo {author} {\bibfnamefont {M.~M.}\ \bibnamefont
  {Desai}}\ and\ \bibinfo {author} {\bibfnamefont {D.~S.}\ \bibnamefont
  {Fisher}},\ }\bibfield  {title} {\bibinfo {title} {{Beneficial
  Mutation-Selection Balance and the Effect of Linkage on Positive
  Selection}},\ }\href {https://doi.org/10.1534/genetics.106.067678} {\bibfield
   {journal} {\bibinfo  {journal} {Genetics}\ }\textbf {\bibinfo {volume}
  {176}},\ \bibinfo {pages} {1759} (\bibinfo {year} {2007})}\BibitemShut
  {NoStop}%
\bibitem [{\citenamefont {Sniegowski}\ and\ \citenamefont
  {Gerrish}(2010)}]{sniegowski2010beneficial}%
  \BibitemOpen
  \bibfield  {author} {\bibinfo {author} {\bibfnamefont {P.~D.}\ \bibnamefont
  {Sniegowski}}\ and\ \bibinfo {author} {\bibfnamefont {P.~J.}\ \bibnamefont
  {Gerrish}},\ }\bibfield  {title} {\bibinfo {title} {Beneficial mutations and
  the dynamics of adaptation in asexual populations},\ }\href
  {https://doi.org/10.1098/rstb.2009.0290} {\bibfield  {journal} {\bibinfo
  {journal} {Philosophical Transactions of the Royal Society B: Biological
  Sciences}\ }\textbf {\bibinfo {volume} {365}},\ \bibinfo {pages} {1255}
  (\bibinfo {year} {2010})}\BibitemShut {NoStop}%
\bibitem [{\citenamefont {Levy}\ \emph {et~al.}(2015)\citenamefont {Levy},
  \citenamefont {Blundell}, \citenamefont {Venkataram}, \citenamefont {Petrov},
  \citenamefont {Fisher},\ and\ \citenamefont
  {Sherlock}}]{levy2015quantitative}%
  \BibitemOpen
  \bibfield  {author} {\bibinfo {author} {\bibfnamefont {S.~F.}\ \bibnamefont
  {Levy}}, \bibinfo {author} {\bibfnamefont {J.~R.}\ \bibnamefont {Blundell}},
  \bibinfo {author} {\bibfnamefont {S.}~\bibnamefont {Venkataram}}, \bibinfo
  {author} {\bibfnamefont {D.~A.}\ \bibnamefont {Petrov}}, \bibinfo {author}
  {\bibfnamefont {D.~S.}\ \bibnamefont {Fisher}},\ and\ \bibinfo {author}
  {\bibfnamefont {G.}~\bibnamefont {Sherlock}},\ }\bibfield  {title} {\bibinfo
  {title} {Quantitative evolutionary dynamics using high-resolution lineage
  tracking},\ }\href {https://doi.org/10.1038/nature14279} {\bibfield
  {journal} {\bibinfo  {journal} {Nature}\ }\textbf {\bibinfo {volume} {519}},\
  \bibinfo {pages} {181} (\bibinfo {year} {2015})}\BibitemShut {NoStop}%
\bibitem [{\citenamefont {Baym}\ \emph {et~al.}(2016)\citenamefont {Baym},
  \citenamefont {Lieberman}, \citenamefont {Kelsic}, \citenamefont {Chait},
  \citenamefont {Gross}, \citenamefont {Yelin},\ and\ \citenamefont
  {Kishony}}]{MEGAplate}%
  \BibitemOpen
  \bibfield  {author} {\bibinfo {author} {\bibfnamefont {M.}~\bibnamefont
  {Baym}}, \bibinfo {author} {\bibfnamefont {T.~D.}\ \bibnamefont {Lieberman}},
  \bibinfo {author} {\bibfnamefont {E.~D.}\ \bibnamefont {Kelsic}}, \bibinfo
  {author} {\bibfnamefont {R.}~\bibnamefont {Chait}}, \bibinfo {author}
  {\bibfnamefont {R.}~\bibnamefont {Gross}}, \bibinfo {author} {\bibfnamefont
  {I.}~\bibnamefont {Yelin}},\ and\ \bibinfo {author} {\bibfnamefont
  {R.}~\bibnamefont {Kishony}},\ }\bibfield  {title} {\bibinfo {title}
  {Spatiotemporal microbial evolution on antibiotic landscapes},\ }\href
  {https://doi.org/10.1126/science.aag0822} {\bibfield  {journal} {\bibinfo
  {journal} {Science}\ }\textbf {\bibinfo {volume} {353}},\ \bibinfo {pages}
  {1147} (\bibinfo {year} {2016})}\BibitemShut {NoStop}%
\bibitem [{CAR(2023)}]{CARPENTER2023e00227}%
  \BibitemOpen
  \bibfield  {title} {\bibinfo {title} {{Have you tried turning it off and on
  again? Oscillating selection to enhance fitness-landscape traversal in
  adaptive laboratory evolution experiments}},\ }\href
  {https://doi.org/https://doi.org/10.1016/j.mec.2023.e00227} {\bibfield
  {journal} {\bibinfo  {journal} {Metabolic Engineering Communications}\
  }\textbf {\bibinfo {volume} {17}},\ \bibinfo {pages} {e00227} (\bibinfo
  {year} {2023})}\BibitemShut {NoStop}%
\bibitem [{\citenamefont {Franco}\ \emph {et~al.}(2020)\citenamefont {Franco},
  \citenamefont {Defeo}, \citenamefont {Piola}, \citenamefont {Barreiro},
  \citenamefont {Yang}, \citenamefont {Ortega}, \citenamefont {Gianelli},
  \citenamefont {Castello}, \citenamefont {Vera}, \citenamefont {Buratti} \emph
  {et~al.}}]{franco2020climate}%
  \BibitemOpen
  \bibfield  {author} {\bibinfo {author} {\bibfnamefont {B.~C.}\ \bibnamefont
  {Franco}}, \bibinfo {author} {\bibfnamefont {O.}~\bibnamefont {Defeo}},
  \bibinfo {author} {\bibfnamefont {A.~R.}\ \bibnamefont {Piola}}, \bibinfo
  {author} {\bibfnamefont {M.}~\bibnamefont {Barreiro}}, \bibinfo {author}
  {\bibfnamefont {H.}~\bibnamefont {Yang}}, \bibinfo {author} {\bibfnamefont
  {L.}~\bibnamefont {Ortega}}, \bibinfo {author} {\bibfnamefont
  {I.}~\bibnamefont {Gianelli}}, \bibinfo {author} {\bibfnamefont {J.~P.}\
  \bibnamefont {Castello}}, \bibinfo {author} {\bibfnamefont {C.}~\bibnamefont
  {Vera}}, \bibinfo {author} {\bibfnamefont {C.}~\bibnamefont {Buratti}}, \emph
  {et~al.},\ }\bibfield  {title} {\bibinfo {title} {{Climate change impacts on
  the atmospheric circulation, ocean, and fisheries in the southwest South
  Atlantic Ocean: a review}},\ }\href
  {https://doi.org/10.1007/s10584-020-02783-6} {\bibfield  {journal} {\bibinfo
  {journal} {Climatic Change}\ }\textbf {\bibinfo {volume} {162}},\ \bibinfo
  {pages} {2359} (\bibinfo {year} {2020})}\BibitemShut {NoStop}%
\bibitem [{\citenamefont {Trenberth}\ \emph {et~al.}(2015)\citenamefont
  {Trenberth}, \citenamefont {Fasullo},\ and\ \citenamefont
  {Shepherd}}]{trenberth2015attribution}%
  \BibitemOpen
  \bibfield  {author} {\bibinfo {author} {\bibfnamefont {K.~E.}\ \bibnamefont
  {Trenberth}}, \bibinfo {author} {\bibfnamefont {J.~T.}\ \bibnamefont
  {Fasullo}},\ and\ \bibinfo {author} {\bibfnamefont {T.~G.}\ \bibnamefont
  {Shepherd}},\ }\bibfield  {title} {\bibinfo {title} {Attribution of climate
  extreme events},\ }\href {https://doi.org/10.1038/nclimate2657} {\bibfield
  {journal} {\bibinfo  {journal} {Nature Climate Change}\ }\textbf {\bibinfo
  {volume} {5}},\ \bibinfo {pages} {725} (\bibinfo {year} {2015})}\BibitemShut
  {NoStop}%
\bibitem [{\citenamefont {Poloczanska}\ \emph {et~al.}(2016)\citenamefont
  {Poloczanska}, \citenamefont {Burrows}, \citenamefont {Brown}, \citenamefont
  {Garc{\'\i}a~Molinos}, \citenamefont {Halpern}, \citenamefont
  {Hoegh-Guldberg}, \citenamefont {Kappel}, \citenamefont {Moore},
  \citenamefont {Richardson}, \citenamefont {Schoeman} \emph
  {et~al.}}]{poloczanska2016responses}%
  \BibitemOpen
  \bibfield  {author} {\bibinfo {author} {\bibfnamefont {E.~S.}\ \bibnamefont
  {Poloczanska}}, \bibinfo {author} {\bibfnamefont {M.~T.}\ \bibnamefont
  {Burrows}}, \bibinfo {author} {\bibfnamefont {C.~J.}\ \bibnamefont {Brown}},
  \bibinfo {author} {\bibfnamefont {J.}~\bibnamefont {Garc{\'\i}a~Molinos}},
  \bibinfo {author} {\bibfnamefont {B.~S.}\ \bibnamefont {Halpern}}, \bibinfo
  {author} {\bibfnamefont {O.}~\bibnamefont {Hoegh-Guldberg}}, \bibinfo
  {author} {\bibfnamefont {C.~V.}\ \bibnamefont {Kappel}}, \bibinfo {author}
  {\bibfnamefont {P.~J.}\ \bibnamefont {Moore}}, \bibinfo {author}
  {\bibfnamefont {A.~J.}\ \bibnamefont {Richardson}}, \bibinfo {author}
  {\bibfnamefont {D.~S.}\ \bibnamefont {Schoeman}}, \emph {et~al.},\ }\bibfield
   {title} {\bibinfo {title} {Responses of marine organisms to climate change
  across oceans},\ }\href {https://doi.org/10.3389/fmars.2016.00062} {\bibfield
   {journal} {\bibinfo  {journal} {Frontiers in Marine Science}\ }\textbf
  {\bibinfo {volume} {3}},\ \bibinfo {pages} {180581} (\bibinfo {year}
  {2016})}\BibitemShut {NoStop}%
\bibitem [{\citenamefont {Pecl}\ \emph {et~al.}(2017)\citenamefont {Pecl} \emph
  {et~al.}}]{biodivredistribition2017}%
  \BibitemOpen
  \bibfield  {author} {\bibinfo {author} {\bibfnamefont {G.~T.}\ \bibnamefont
  {Pecl}} \emph {et~al.},\ }\bibfield  {title} {\bibinfo {title} {Biodiversity
  redistribution under climate change: Impacts on ecosystems and human
  well-being},\ }\href {https://doi.org/10.1126/science.aai9214} {\bibfield
  {journal} {\bibinfo  {journal} {Science}\ }\textbf {\bibinfo {volume}
  {355}},\ \bibinfo {pages} {eaai9214} (\bibinfo {year} {2017})}\BibitemShut
  {NoStop}%
\bibitem [{\citenamefont {Gong}\ and\ \citenamefont
  {Hamazaki}(2022)}]{gong2022bounds}%
  \BibitemOpen
  \bibfield  {author} {\bibinfo {author} {\bibfnamefont {Z.}~\bibnamefont
  {Gong}}\ and\ \bibinfo {author} {\bibfnamefont {R.}~\bibnamefont
  {Hamazaki}},\ }\href {https://doi.org/10.1142/S0217979222300079} {\bibinfo
  {title} {Bounds in nonequilibrium quantum dynamics}} (\bibinfo {year}
  {2022})\BibitemShut {NoStop}%
\bibitem [{\citenamefont {Carabba}\ \emph {et~al.}(2022)\citenamefont
  {Carabba}, \citenamefont {H{\"o}rnedal},\ and\ \citenamefont {del
  Campo}}]{AdC2022Quantum}%
  \BibitemOpen
  \bibfield  {author} {\bibinfo {author} {\bibfnamefont {N.}~\bibnamefont
  {Carabba}}, \bibinfo {author} {\bibfnamefont {N.}~\bibnamefont
  {H{\"o}rnedal}},\ and\ \bibinfo {author} {\bibfnamefont {A.}~\bibnamefont
  {del Campo}},\ }\bibfield  {title} {\bibinfo {title} {Quantum speed limits on
  operator flows and correlation functions},\ }\href
  {https://doi.org/10.22331/q-2022-12-22-884} {\bibfield  {journal} {\bibinfo
  {journal} {Quantum}\ }\textbf {\bibinfo {volume} {6}},\ \bibinfo {pages}
  {884} (\bibinfo {year} {2022})}\BibitemShut {NoStop}%
\bibitem [{\citenamefont {Uffink}\ and\ \citenamefont
  {Van~Lith}(1999)}]{uffink1999thermodynamic}%
  \BibitemOpen
  \bibfield  {author} {\bibinfo {author} {\bibfnamefont {J.}~\bibnamefont
  {Uffink}}\ and\ \bibinfo {author} {\bibfnamefont {J.}~\bibnamefont
  {Van~Lith}},\ }\bibfield  {title} {\bibinfo {title} {Thermodynamic
  uncertainty relations},\ }\href {https://doi.org/10.1023/A:1018811305766}
  {\bibfield  {journal} {\bibinfo  {journal} {Foundations of physics}\ }\textbf
  {\bibinfo {volume} {29}},\ \bibinfo {pages} {655} (\bibinfo {year}
  {1999})}\BibitemShut {NoStop}%
\bibitem [{\citenamefont {Barato}\ and\ \citenamefont
  {Seifert}(2015)}]{PhysRevLett.114.158101}%
  \BibitemOpen
  \bibfield  {author} {\bibinfo {author} {\bibfnamefont {A.~C.}\ \bibnamefont
  {Barato}}\ and\ \bibinfo {author} {\bibfnamefont {U.}~\bibnamefont
  {Seifert}},\ }\bibfield  {title} {\bibinfo {title} {Thermodynamic uncertainty
  relation for biomolecular processes},\ }\href
  {https://doi.org/10.1103/PhysRevLett.114.158101} {\bibfield  {journal}
  {\bibinfo  {journal} {Phys. Rev. Lett.}\ }\textbf {\bibinfo {volume} {114}},\
  \bibinfo {pages} {158101} (\bibinfo {year} {2015})}\BibitemShut {NoStop}%
\bibitem [{\citenamefont {Vo}\ \emph {et~al.}(2020)\citenamefont {Vo},
  \citenamefont {Van~Vu},\ and\ \citenamefont
  {Hasegawa}}]{PhysRevE.102.062132}%
  \BibitemOpen
  \bibfield  {author} {\bibinfo {author} {\bibfnamefont {V.~T.}\ \bibnamefont
  {Vo}}, \bibinfo {author} {\bibfnamefont {T.}~\bibnamefont {Van~Vu}},\ and\
  \bibinfo {author} {\bibfnamefont {Y.}~\bibnamefont {Hasegawa}},\ }\bibfield
  {title} {\bibinfo {title} {Unified approach to classical speed limit and
  thermodynamic uncertainty relation},\ }\href
  {https://doi.org/10.1103/PhysRevE.102.062132} {\bibfield  {journal} {\bibinfo
   {journal} {Phys. Rev. E}\ }\textbf {\bibinfo {volume} {102}},\ \bibinfo
  {pages} {062132} (\bibinfo {year} {2020})}\BibitemShut {NoStop}%
\bibitem [{\citenamefont {Horowitz}\ and\ \citenamefont
  {Gingrich}(2020)}]{horowitz2020thermodynamic}%
  \BibitemOpen
  \bibfield  {author} {\bibinfo {author} {\bibfnamefont {J.~M.}\ \bibnamefont
  {Horowitz}}\ and\ \bibinfo {author} {\bibfnamefont {T.~R.}\ \bibnamefont
  {Gingrich}},\ }\bibfield  {title} {\bibinfo {title} {Thermodynamic
  uncertainty relations constrain non-equilibrium fluctuations},\ }\href
  {https://doi.org/s41567-019-0702-6} {\bibfield  {journal} {\bibinfo
  {journal} {Nature Physics}\ }\textbf {\bibinfo {volume} {16}},\ \bibinfo
  {pages} {15} (\bibinfo {year} {2020})}\BibitemShut {NoStop}%
\bibitem [{\citenamefont {Dechant}\ and\ \citenamefont
  {Sasa}(2021)}]{PhysRevX.11.041061}%
  \BibitemOpen
  \bibfield  {author} {\bibinfo {author} {\bibfnamefont {A.}~\bibnamefont
  {Dechant}}\ and\ \bibinfo {author} {\bibfnamefont {S.-i.}\ \bibnamefont
  {Sasa}},\ }\bibfield  {title} {\bibinfo {title} {Improving thermodynamic
  bounds using correlations},\ }\href
  {https://doi.org/10.1103/PhysRevX.11.041061} {\bibfield  {journal} {\bibinfo
  {journal} {Phys. Rev. X}\ }\textbf {\bibinfo {volume} {11}},\ \bibinfo
  {pages} {041061} (\bibinfo {year} {2021})}\BibitemShut {NoStop}%
\bibitem [{\citenamefont {Das}\ and\ \citenamefont
  {Green}(2023)}]{JasonGreenPRR2023}%
  \BibitemOpen
  \bibfield  {author} {\bibinfo {author} {\bibfnamefont {S.}~\bibnamefont
  {Das}}\ and\ \bibinfo {author} {\bibfnamefont {J.~R.}\ \bibnamefont
  {Green}},\ }\bibfield  {title} {\bibinfo {title} {Speed limits on
  deterministic chaos and dissipation},\ }\href
  {https://doi.org/10.1103/PhysRevResearch.5.L012016} {\bibfield  {journal}
  {\bibinfo  {journal} {Phys. Rev. Res.}\ }\textbf {\bibinfo {volume} {5}},\
  \bibinfo {pages} {L012016} (\bibinfo {year} {2023})}\BibitemShut {NoStop}%
\bibitem [{\citenamefont {Garc\'{\i}a-Pintos}\ \emph
  {et~al.}(2023)\citenamefont {Garc\'{\i}a-Pintos}, \citenamefont {Brady},
  \citenamefont {Bringewatt},\ and\ \citenamefont {Liu}}]{QSLAnneal}%
  \BibitemOpen
  \bibfield  {author} {\bibinfo {author} {\bibfnamefont {L.~P.}\ \bibnamefont
  {Garc\'{\i}a-Pintos}}, \bibinfo {author} {\bibfnamefont {L.~T.}\ \bibnamefont
  {Brady}}, \bibinfo {author} {\bibfnamefont {J.}~\bibnamefont {Bringewatt}},\
  and\ \bibinfo {author} {\bibfnamefont {Y.-K.}\ \bibnamefont {Liu}},\
  }\bibfield  {title} {\bibinfo {title} {Lower bounds on quantum annealing
  times},\ }\href {https://doi.org/10.1103/PhysRevLett.130.140601} {\bibfield
  {journal} {\bibinfo  {journal} {Phys. Rev. Lett.}\ }\textbf {\bibinfo
  {volume} {130}},\ \bibinfo {pages} {140601} (\bibinfo {year}
  {2023})}\BibitemShut {NoStop}%
\bibitem [{\citenamefont {Crow}\ and\ \citenamefont
  {Morton}(1955)}]{CrowGeneticDrift1955}%
  \BibitemOpen
  \bibfield  {author} {\bibinfo {author} {\bibfnamefont {J.~F.}\ \bibnamefont
  {Crow}}\ and\ \bibinfo {author} {\bibfnamefont {N.~E.}\ \bibnamefont
  {Morton}},\ }\bibfield  {title} {\bibinfo {title} {Measurement of gene
  frequency drift in small populations},\ }\href
  {https://doi.org/10.2307/2405589} {\bibfield  {journal} {\bibinfo  {journal}
  {Evolution}\ }\textbf {\bibinfo {volume} {9}},\ \bibinfo {pages} {202}
  (\bibinfo {year} {1955})}\BibitemShut {NoStop}%
\bibitem [{\citenamefont {Traulsen}\ and\ \citenamefont
  {Hauert}(2009)}]{TraulsenReview2009}%
  \BibitemOpen
  \bibfield  {author} {\bibinfo {author} {\bibfnamefont {A.}~\bibnamefont
  {Traulsen}}\ and\ \bibinfo {author} {\bibfnamefont {C.}~\bibnamefont
  {Hauert}},\ }\bibinfo {title} {Stochastic evolutionary game dynamics},\ in\
  \href {https://doi.org/https://doi.org/10.1002/9783527628001.ch2} {\emph
  {\bibinfo {booktitle} {Reviews of Nonlinear Dynamics and Complexity}}}\
  (\bibinfo  {publisher} {John Wiley \& Sons, Ltd},\ \bibinfo {year} {2009})\
  Chap.~\bibinfo {chapter} {2}, pp.\ \bibinfo {pages} {25--61}\BibitemShut
  {NoStop}%
\bibitem [{\citenamefont {Tataru}\ \emph {et~al.}(2016)\citenamefont {Tataru},
  \citenamefont {Simonsen}, \citenamefont {Bataillon},\ and\ \citenamefont
  {Hobolth}}]{WrightFisher2016}%
  \BibitemOpen
  \bibfield  {author} {\bibinfo {author} {\bibfnamefont {P.}~\bibnamefont
  {Tataru}}, \bibinfo {author} {\bibfnamefont {M.}~\bibnamefont {Simonsen}},
  \bibinfo {author} {\bibfnamefont {T.}~\bibnamefont {Bataillon}},\ and\
  \bibinfo {author} {\bibfnamefont {A.}~\bibnamefont {Hobolth}},\ }\bibfield
  {title} {\bibinfo {title} {{Statistical Inference in the {Wright-Fisher}
  Model Using Allele Frequency Data}},\ }\href
  {https://doi.org/10.1093/sysbio/syw056} {\bibfield  {journal} {\bibinfo
  {journal} {Systematic Biology}\ }\textbf {\bibinfo {volume} {66}},\ \bibinfo
  {pages} {e30} (\bibinfo {year} {2016})}\BibitemShut {NoStop}%
\bibitem [{\citenamefont {Huang}\ \emph {et~al.}(2015)\citenamefont {Huang},
  \citenamefont {Hauert},\ and\ \citenamefont {Traulsen}}]{TraulsenPNAS2015}%
  \BibitemOpen
  \bibfield  {author} {\bibinfo {author} {\bibfnamefont {W.}~\bibnamefont
  {Huang}}, \bibinfo {author} {\bibfnamefont {C.}~\bibnamefont {Hauert}},\ and\
  \bibinfo {author} {\bibfnamefont {A.}~\bibnamefont {Traulsen}},\ }\bibfield
  {title} {\bibinfo {title} {Stochastic game dynamics under demographic
  fluctuations},\ }\href {https://doi.org/10.1073/pnas.1418745112} {\bibfield
  {journal} {\bibinfo  {journal} {Proceedings of the National Academy of
  Sciences}\ }\textbf {\bibinfo {volume} {112}},\ \bibinfo {pages} {9064}
  (\bibinfo {year} {2015})}\BibitemShut {NoStop}%
\bibitem [{\citenamefont {Vasconcelos}\ \emph {et~al.}(2017)\citenamefont
  {Vasconcelos}, \citenamefont {Santos}, \citenamefont {Santos},\ and\
  \citenamefont {Pacheco}}]{VasconcelosPRL2017}%
  \BibitemOpen
  \bibfield  {author} {\bibinfo {author} {\bibfnamefont {V.~V.}\ \bibnamefont
  {Vasconcelos}}, \bibinfo {author} {\bibfnamefont {F.~P.}\ \bibnamefont
  {Santos}}, \bibinfo {author} {\bibfnamefont {F.~C.}\ \bibnamefont {Santos}},\
  and\ \bibinfo {author} {\bibfnamefont {J.~M.}\ \bibnamefont {Pacheco}},\
  }\bibfield  {title} {\bibinfo {title} {Stochastic dynamics through
  hierarchically embedded markov chains},\ }\href
  {https://doi.org/10.1103/PhysRevLett.118.058301} {\bibfield  {journal}
  {\bibinfo  {journal} {Phys. Rev. Lett.}\ }\textbf {\bibinfo {volume} {118}},\
  \bibinfo {pages} {058301} (\bibinfo {year} {2017})}\BibitemShut {NoStop}%
\bibitem [{\citenamefont {Berg}\ and\ \citenamefont
  {Ellers}(2010)}]{berg2010plasticity}%
  \BibitemOpen
  \bibfield  {author} {\bibinfo {author} {\bibfnamefont {M.~P.}\ \bibnamefont
  {Berg}}\ and\ \bibinfo {author} {\bibfnamefont {J.}~\bibnamefont {Ellers}},\
  }\bibfield  {title} {\bibinfo {title} {Trait plasticity in species
  interactions: a driving force of community dynamics},\ }\href
  {https://doi.org/10.1007/s10682-009-9347-8} {\bibfield  {journal} {\bibinfo
  {journal} {Evolutionary Ecology}\ }\textbf {\bibinfo {volume} {24}},\
  \bibinfo {pages} {617} (\bibinfo {year} {2010})}\BibitemShut {NoStop}%
\bibitem [{\citenamefont {Whitley}(1994)}]{whitley1994genetic}%
  \BibitemOpen
  \bibfield  {author} {\bibinfo {author} {\bibfnamefont {D.}~\bibnamefont
  {Whitley}},\ }\bibfield  {title} {\bibinfo {title} {A genetic algorithm
  tutorial},\ }\href {https://doi.org/10.1007/BF00175354} {\bibfield  {journal}
  {\bibinfo  {journal} {Statistics and computing}\ }\textbf {\bibinfo {volume}
  {4}},\ \bibinfo {pages} {65} (\bibinfo {year} {1994})}\BibitemShut {NoStop}%
\bibitem [{\citenamefont {Yu}\ and\ \citenamefont
  {Gen}(2010)}]{EvolutionaryAlgo2010introduction}%
  \BibitemOpen
  \bibfield  {author} {\bibinfo {author} {\bibfnamefont {X.}~\bibnamefont
  {Yu}}\ and\ \bibinfo {author} {\bibfnamefont {M.}~\bibnamefont {Gen}},\
  }\href {https://doi.org/10.1007/978-1-84996-129-5} {\emph {\bibinfo {title}
  {Introduction to evolutionary algorithms}}}\ (\bibinfo  {publisher} {Springer
  Science \& Business Media},\ \bibinfo {year} {2010})\BibitemShut {NoStop}%
\bibitem [{\citenamefont {Hodgson}\ and\ \citenamefont
  {Knudsen}(2010)}]{HODGSON201012}%
  \BibitemOpen
  \bibfield  {author} {\bibinfo {author} {\bibfnamefont {G.~M.}\ \bibnamefont
  {Hodgson}}\ and\ \bibinfo {author} {\bibfnamefont {T.}~\bibnamefont
  {Knudsen}},\ }\bibfield  {title} {\bibinfo {title} {Generative replication
  and the evolution of complexity},\ }\href
  {https://doi.org/https://doi.org/10.1016/j.jebo.2010.03.008} {\bibfield
  {journal} {\bibinfo  {journal} {Journal of Economic Behavior \&
  Organization}\ }\textbf {\bibinfo {volume} {75}},\ \bibinfo {pages} {12}
  (\bibinfo {year} {2010})},\ \bibinfo {note} {transdisciplinary Perspectives
  on Economic Complexity}\BibitemShut {NoStop}%
\end{thebibliography}%

\widetext
\clearpage

\appendix

 \part*{ 
 \begin{center}
 \normalsize{
 APPENDIX 
 } 
 \end{center}
 }

This Supplementary Material includes detailed proofs of Eqs.~(1),~(2),~(9),~(11),~(14), and~(15) in the main text. I also show that the techniques developed in this article can be applied to other dynamical models that incorporate stochastic forces, and illustrate the rate limits for the replicator-mutator equation on a toy model.

\section{Rate limits for arbitrary evolutionary dynamics}
\label{sec-app:GeneralRateLimits}
 I prove Eqs.~(1) and~(2) in the main text. I also show that the variance in the growth rate of a population equals the classical Fisher information of the frequency distribution of the population.

\vspace{7pt}

Let $n_j$ be the time-dependent population of type $j$ and $N \coloneqq \sum_j n_j$, so that $\{p_j \coloneqq n_j/N\}$ is a normalized distribution. The rate of change of the average $\langle A \rangle \coloneqq \sum_j \frac{n_j}{N} a_j$ over the population satisfies
\begin{align}
    \dt \langle A \rangle - \langle \dot A \rangle &= \sum_j a_j \frac{\dot n_j}{N} - \frac{\dot N}{ N} \sum_j a_j \frac{n_j}{N}  = \sum_j a_j \frac{n_j}{N} \frac{\dot n_j}{n_j} - \frac{\dot N}{ N} \langle A \rangle 
    = \sum_j \delta a_j \frac{n_j}{N} \frac{\dot n_j}{n_j} + \langle A \rangle \sum_j  \frac{n_j}{N} \frac{\dot n_j}{n_j} - \frac{\dot N}{N} \langle A \rangle \nonumber \\
    &= \sum_j \delta a_j \frac{n_j}{N} \frac{\dot n_j}{n_j} = \sum_j \delta a_j \frac{n_j}{N} r_j = \sum_j \delta a_j \frac{n_j}{N} \delta r_j,
\end{align}
where I denote $\delta a_j \coloneqq a_j - \langle A \rangle$ and $\delta r_j \coloneqq r_j - \langle r \rangle$, with $r_j \coloneqq \frac{\dot n_j}{n_j}$. In the last step I used that $\langle r \rangle \sum_j \delta a_j n_j/N = 0$ given that $\langle \delta A \rangle = 0$. 

That is, 
\begin{align}
    \dt \langle A \rangle - \langle \dot A \rangle &= \sum_j \delta a_j \, \delta r_j \frac{n_j}{N} = \cov(r,A),
\end{align}
where the covariance is defined by $\cov(r,A) \coloneqq \langle \delta A \, \delta r \rangle$. This proves Eq.~(1) in the main text. 

Using that the covariance of two quantities is bounded by the product of their standard deviations, I obtain that
\begin{align}
\label{eq-app:speedlimit}
    \left| \dt \langle A \rangle  - \langle \dot A \rangle \right| &= \left| \cov(r,A) \right| \leq \sigma_r \, \sigma_A,
\end{align}
where $\sigma_A \coloneqq \sqrt{ \langle A^2 \rangle - \langle A \rangle^2 }$ denotes the standard deviation of $A$.
Equation~\eqref{eq-app:speedlimit} implies that, for any trait $A$, the evolution of the mean is fast only if $A$ or $r$ have variability -- i.e., their standard deviations $\sigma_r$ and $\sigma_A$ cannot be small for fast evolution. This proves Eq.~(2) in the main text.

Defining the normalized frequency distribution $p_j \coloneqq n_j/N$ and using that $\dot p_j \nobreak = \nobreak \dot n_j/N \nobreak-\nobreak n_j \dot N/N^2$,
it holds that 
\begin{align}
\label{eq-app:ntop}
\frac{\dot p_j}{p_j} = \frac{\dot n_j}{n_j} - \frac{\dot N}{N}.
\end{align}
 Then, 
\begin{align}
\info &\coloneqq \sum_j p_j \bigg( \frac{\dot p_j}{p_j} \bigg)^2 = \sum_j \frac{n_j}{N} \Bigg( \frac{\dot n_j}{n_j} - \frac{\dot N}{N} \Bigg)^2 = \sum_j \frac{n_j}{N} \bigg( \frac{\dot n_j}{n_j}\bigg)^2 - 2\sum_j \frac{n_j}{N} \frac{\dot n_j}{n_j} \frac{\dot N}{N} + \sum_j \frac{n_j}{N} \frac{\dot N^2}{N^2} = \sum_j \frac{n_j}{N} \bigg( \frac{\dot n_j}{n_j}\bigg)^2 - \Bigg(\frac{\dot N}{N} \Bigg)^2 \nonumber \\
&= \sum_j \frac{n_j}{N} \bigg( \frac{\dot n_j}{n_j}\bigg)^2 - \Bigg( \sum_j \frac{n_j}{N}\frac{\dot n_j}{n_j} \Bigg)^2 = \sum_j \frac{n_j}{N} r_j^2 - \Bigg( \sum_j \frac{n_j}{N} r_j \Bigg)^2 = \left\langle r^2 \right\rangle - \langle r \rangle^2 \equiv \left(\sigma_r \right)^2.
\end{align}
That is, the variance in the rate $r$ equals the Fisher information of the normalized distribution $p_j \coloneqq n_j/N$.
Note, too, that Eq.~(\ref{eq-app:ntop}) implies that $\cov(A,r) = \cov(A,f)$ for dynamics governed by the replicator equation.

\clearpage
\section{Limits to evolutionary processes with mutations}
\label{sec-app:limitmutations}
 In this section I prove Equations~(9) and~(11) in the main text.

\vspace{7pt}

Let $\delta_{\Pi} a_j \coloneqq a_j - \langle A \rangle_\Pi$ and $\delta f_j \coloneqq f_j - \langle f \rangle$, where $\langle A \rangle_\Pi \coloneqq \sum_j \Pi_j a_j$ and $\langle f \rangle \coloneqq \sum_j p_j f_j$ are means with respect to the distribution $\Pi$ with components $\Pi \coloneqq \sum_k p_k Q_{kj}$ and the distribution $p$, respectively.
Then, using conservation of probability (which implies $\sum_j \dot p_j \langle A \rangle_\Pi = 0$) and the mutator-replicator equation, it holds that
\begin{align}
\label{eq-aux:rate}
\frac{d\langle A \rangle}{dt} - \langle \dot A \rangle &= \sum_j \dot p_j a_j = \sum_j \dot p_j \delta_\Pi a_j  \nonumber \\
&= \sum_{jk} Q_{kj} f_ k p_k  \delta_\Pi a_j - \sum_j p_j \delta_\Pi a_j \langle f \rangle \nonumber \\
&= \sum_{jk} Q_{kj} \delta\! f_ k \, p_k \, \delta_\Pi a_j + \sum_{jk} Q_{kj} \langle f \rangle \, p_k \, \delta_\Pi a_j - \langle f \rangle \langle A \rangle + \langle f \rangle \langle A \rangle_\Pi  \nonumber \\
&= \sum_{jk} Q_{kj} \delta\! f_ k \, p_k \, \delta_\Pi a_j +   \langle f \rangle  \sum_{j} \Pi_j \, \delta_\Pi a_j - \langle f \rangle \langle A \rangle + \langle f \rangle \langle A \rangle_\Pi  \nonumber \\
&= \sum_{jk} Q_{kj} \delta\! f_ k \, p_k \, \delta_\Pi a_j  - \langle f \rangle \big( \langle A \rangle - \langle A \rangle_\Pi \big). 
\end{align}

The Cauchy-Schwarz inequality says that $\sum_{\alpha} X_\alpha Y_\alpha \leq \sqrt{\left( \sum_\alpha X_\alpha^2 \right) \left( \sum_\alpha Y_\alpha^2 \right)}$. Applying it to the first term in the last line, with $X_\alpha = \sqrt{Q_{kj}} \delta f_k \sqrt{p_k}$ and $Y_\alpha = \sqrt{Q_{kj} p_k} \delta_\Pi a_j$ where $\alpha$ denotes both indexes $\{ j,k \}$, gives
\begin{align}
\label{eq-aux:bound}
\Bigg( \sum_{jk} Q_{kj} \delta\! f_ k \, p_k \, \delta_\Pi a_j \Bigg)^2 &\leq \Bigg( \sum_{jk} Q_{kj} \left( \delta\! f_ k \right)^2 \, p_k  \Bigg) \Bigg( \sum_{jk} Q_{kj}  p_k \, (\delta_\Pi a_j)^2 \Bigg) \nonumber \\
&= \Bigg( \sum_{k} \left( \delta\! f_ k \right)^2 \, p_k  \Bigg) \Bigg( \sum_{j} \Pi_k \, (\delta_\Pi a_j)^2 \Bigg) = \left(\sigma_f\right)^2 \, \left( \sigma^\Pi_A\right)^2,
\end{align}
where I used that $\sum_j Q_{kj} = 1$. Combining Eqs.~\eqref{eq-aux:rate} and~\eqref{eq-aux:bound} leads to 
\begin{align}
\label{eq-app:AuxRateLimitMutations}
\left| \frac{d\langle A \rangle}{dt} - \langle \dot A \rangle - \langle f \rangle \Big( \langle A \rangle_\Pi - \langle A \rangle \Big)  \right|  = \left| \sum_{jk} Q_{kj} \delta\! f_ k \, p_k \, \delta_\Pi a_j \right| \leq \sigma^\Pi_A \, \sigma_f,
\end{align}
which proves Equation (9) in the main text.
 
Choosing a trait with components $a_j \equiv I_j = -\ln p_j$ gives:
\begin{align}
    \langle A \rangle &= \sum_j p_j (-\ln p_j) = S\\
    \langle A \rangle_\Pi - \langle A \rangle &= \sum_j \Pi_j (-\ln p_j) - \sum_j p_j (-\ln p_j) = S(p||\Pi).
\end{align}
Using that $\langle \dot A \rangle = \sum_j p_j (-\dot p_j/p_j) = - \sum_j \dot p_j = 0 $ yields Equation~(11) in the main text.

\section{Limits to stochastic evolutionary processes}
\label{sec-app:limitstochastic}
In this section I prove Equation (14) in the main text. 

 \vspace{7pt}

 Consider a stochastic replicator-mutator equation,
\begin{align}
\label{eq-app:stochastic}
d p_j &= \sum_k p_k Q_{kj} \big( f_k - \langle f \rangle  \big) dt   + p_j \left( \gamma_j dW_j - \sum_l \gamma_l p_l dW_l \right), 
\end{align}
where $\gamma_j$ characterizes the strength of the stochastic driving forces for population $j$, and $dW_j$ are Wiener noises, which satisfy $\overline{dW_j dW_k} = \delta_{jk} dt$ and $dW_j^2 = dt$. 

Let us focus on the noise-averaged change in an expectation value over a time $\tau$ relative to the change 
\begin{align}
 \big[d \langle A \rangle \nobreak- \nobreak\langle dA \rangle \big] \Big|_{\textnormal{replicator-mutator}} \nobreak= \sum_j a_j d p_j \Big|_{\textnormal{replicator-mutator}}  =\nobreak \sum_{jk}  p_k Q_{kj} \big( f_k - \langle f \rangle  \big) a_j  dt   
\end{align}
in $\langle A \rangle$ due to state changes from natural selection and mutations, as modeled by the replicator-mutator equation. It is given by
\begin{align}
\label{eq-app:aux1stoch}
\overline{\left| \frac{1}{\tau} \int_0^\tau \bigg( d \langle A \rangle - \langle dA \rangle   - \sum_{jk} p_k Q_{kj} \big( f_k - \langle f \rangle  \big) a_j  dt \bigg) \right|^2} &=  \overline{\left| \frac{1}{\tau}  \int_0^\tau \bigg( \sum_j dp_j a_j   - \sum_{jk} p_k Q_{kj} \big( f_k - \langle f \rangle  \big) \delta a_j dt \bigg) \right|^2} \nonumber \\
&=  \overline{ \left| \frac{1}{\tau} \int_0^\tau \bigg( \sum_{j}  \delta a_j \big( dp_j - \sum_k p_k Q_{kj} \big( f_k - \langle f \rangle  \big) dt \big) \bigg) \right|^2} \nonumber \\
&=  \overline{ \left| \frac{1}{\tau} \int_0^\tau \sum_j  \delta a_j p_j \left( \gamma_j dW_j - \sum_l \gamma_l p_l dW_l \right)  \right|^2} \nonumber \\ 
&= \overline{ \left| \frac{1}{\tau} \int_0^\tau \sum_j \delta a_j p_j Y_j \right|^2 } \nonumber \\
& = \frac{1}{\tau^2} \int_0^\tau \int_0^\tau  \sum_{jk} \overline{  \delta a_j \delta a_k p_j p_k  Y_j Y_k },
\end{align}
where I used that $\delta a_j \coloneqq a_j - \langle a \rangle$, that probability is conserved, $\sum_j dp_j = 0$, and Eq.~(\ref{eq-app:stochastic}). Here, 
\begin{align}
Y_j &\coloneqq \left( \gamma_j dW_j - \sum_l \gamma_l p_l dW_l \right).
\end{align}

 In order to prevent confusion, I explicitly include the time-dependence of the noise terms. For Wiener processes, it holds that $\overline{dW_j^t dW_j^{t'}} = \delta_{jk} \delta_{t t'}$, and that the noise terms are independent from all other functions (in the It\^o picture), so that $\overline{p_j^t dW_j^t} = \overline{p_j^t} \,\, \overline{dW_j^t} = 0$. The integrals that lack a line element $dt$, proportional to a noise term, correspond to It\^o integrals~\cite{jacobs_stochastic}.

Using the rules of It\^o calculus described above and that $\sum_j \delta a_j p_j = 0$ gives that 
 \begin{align}
 \label{eq-app:aux2stoch}
 \int_0^\tau \int_0^\tau  \sum_{jk} \overline{ \delta a_j \delta a_k  p_j p_k Y_j Y_k }   &= \int_0^\tau \int_0^\tau \sum_{jk} \overline{ \delta a_j \delta a_k  p_j p_k \gamma_j \gamma_k dW_j^t dW_k^{t'} } \nonumber \\
 &-2\int_0^\tau \int_0^\tau \sum_{jk} \overline{ \delta a_j \delta a_k  p_j p_k \gamma_j dW_j^t \sum_l p_l \gamma_l  dW_l^{t'} }   \nonumber \\
 &+  \int_0^\tau \int_0^\tau \sum_{jk} \overline{ \delta a_j \delta a_k  p_j p_k \left( \sum_l p_l \gamma_l dW_l^t \right) \left( \sum_m p_m \gamma_m dW_m^{t'} \right) } \nonumber \\
 &=   \int_0^\tau \sum_{j} \overline{ \left(\delta a_j \right)^2 \gamma_j^2 \,  p_j^2 } dt 
 - 2 \int_0^\tau \sum_{jk} \overline{ \delta a_j \delta a_k  p_j p_k \gamma_j  p_j \gamma_j    } dt \nonumber \\
 &+  \int_0^\tau \sum_{jk} \overline{ \delta a_j \delta a_k  p_j p_k \left( \sum_l p_l^2 \gamma_l^2 \right) } dt \nonumber \\
 &= \overline{\int_0^\tau \sum_{j} \left(\delta a_j \right)^2 \gamma_j^2 p_j^2} dt .
\end{align}
 
 Combining Eq.~(\ref{eq-app:aux1stoch}) and~(\ref{eq-app:aux2stoch}) results in
  \begin{align}
  \label{eq-app:auxRateLimitStochastic}
\overline{\left( \frac{1}{\tau} \int_0^\tau d \langle A \rangle - \langle dA \rangle - \sum_{jk} p_k Q_{kj} \big( f_k - \langle f \rangle  \big) a_j  dt \right)^2} &= \frac{1}{\tau^2} \overline{\int_0^\tau \sum_{j} \left(\delta a_j \right)^2 \gamma_j^2 p_j^2} dt \nonumber \\
& \leq  \| \gamma \|_\infty^2 \frac{1}{\tau^2} \overline{\int_0^\tau  \sum_{j} \left(\delta a_j \right)^2 p_j^2 } dt \nonumber \\
& \leq  \| \gamma \|_\infty^2 \frac{1}{\tau^2} \overline{\int_0^\tau  \sum_{j} \left(\delta a_j \right)^2 p_j } dt \nonumber \\
&=  \frac{\| \gamma \|_\infty^2}{\tau} \overline{ \frac{1}{\tau} \int_0^\tau  \left( \sigma_A \right)^2 } dt.
\end{align}
This follows from 
 $\gamma_j \leq \| \gamma \|_\infty \coloneqq \max_j \{ \gamma_j \}$ and $p_j^2 \leq p_j$.
This proves Eq.~(14) in the main text. 

\subsection{Limits under other stochastic dynamics}
 An alternative stochastic replicator equation of the form 
\begin{align}
\label{eq-app:stochreplicator}
d p_j &= p_j \left( f_j - \langle f \rangle -\gamma_j^2 p_j + \sum_l \gamma_l^2 p_l^2 \right) dt   + p_j \left( \gamma_j dW_j - \sum_l \gamma_l p_l dW_l \right)
\end{align} 
was derived in Ref.~\cite{cabrales2000stochastic}. For simplicity, I assume no mutations in this sub-appendix, but it is easy to generalize the results that follow to account for them. Here, $\gamma_j$ characterizes the strength of the stochastic driving forces for population $j$, and $dW_j$ are Wiener noises, which satisfy $\overline{dW_j dW_k} = \delta_{jk} dt$ and $dW_j^2 = dt$.

Consider the noise-averaged change in an expectation value over a time $\tau$,  relative to the change $\big[d \langle A \rangle \nobreak- \nobreak\langle dA \rangle \big] \Big|_{\textnormal{replicator}} \nobreak=\nobreak \cov(A,f)dt$ in $\langle A \rangle$ due state changes from natural selection as modeled by the replicator equation.

Following similar calculations as above, one finds that all cross terms proportional to a single noise term vanish upon averaging, and therefore
 \begin{align}
\overline{\left( \frac{1}{\tau} \int_0^\tau d \langle A \rangle - \langle dA \rangle  -\cov(A,f)dt \right)^2} &=  
 \frac{1}{\tau^2} \int_0^\tau \int_0^\tau \sum_{jk} \overline{ \delta a_j \delta a_k \left( dp_j  - p_j \delta f_j dt \right) \left( dp_k - p_k \delta f_k dt' \right)} \nonumber \\
&= \frac{1}{\tau^2} \int_0^\tau \int_0^\tau  \sum_{jk}\overline{ \delta a_j \delta a_k  \left( p_j X_j - p_j \delta f_j \right) \left( p_k  X_k - p_k \delta f_k \right) } dt dt' \nonumber \\
& + \frac{1}{\tau^2} \int_0^\tau \int_0^\tau  \sum_{jk} \overline{  \delta a_j \delta a_k p_j p_k  Y_j Y_k },
\end{align}
by using that $dp_j = p_j X_j  dt + p_j Y_j$, with the notation
\begin{align}
X_j &\coloneqq \left( f_j - \langle f \rangle -\gamma_j^2 p_j + \sum_l \gamma_l^2 p_l^2 \right)  =  \left( \delta f_j  -\gamma_j^2 p_j + \sum_l \gamma_l^2 p_l^2 \right)\\
Y_j &\coloneqq \left( \gamma_j dW_j - \sum_l \gamma_l p_l dW_l \right).
\end{align}

Then, using the rules of It\^o calculus described above and that $\sum_j \delta a_j p_j = 0$ gives that the second term is
 \begin{align}
 \int_0^\tau \int_0^\tau  \sum_{jk} \overline{ \delta a_j \delta a_k  p_j p_k Y_j Y_k }   &= \int_0^\tau \int_0^\tau \sum_{jk} \overline{ \delta a_j \delta a_k  p_j p_k \gamma_j \gamma_k dW_j^t dW_k^{t'} } \nonumber \\
 &-2\int_0^\tau \int_0^\tau \sum_{jk} \overline{ \delta a_j \delta a_k  p_j p_k \gamma_j dW_j^t \sum_l p_l \gamma_l  dW_l^{t'} }   \nonumber \\
 &+  \int_0^\tau \int_0^\tau \sum_{jk} \overline{ \delta a_j \delta a_k  p_j p_k \left( \sum_l p_l \gamma_l dW_l^t \right) \left( \sum_m p_m \gamma_m dW_m^{t'} \right) } \nonumber \\
 &=   \int_0^\tau \sum_{j} \overline{ \left(\delta a_j \right)^2 \gamma_j^2 \,  p_j^2 } dt 
 - 2 \int_0^\tau \sum_{jk} \overline{ \delta a_j \delta a_k  p_j p_k \gamma_j  p_j \gamma_j    } dt \nonumber \\
 &+  \int_0^\tau \sum_{jk} \overline{ \delta a_j \delta a_k  p_j p_k \left( \sum_l p_l^2 \gamma_l^2 \right) } dt \nonumber \\
 &= \overline{\int_0^\tau \sum_{j} \left(\delta a_j \right)^2 \gamma_j^2 p_j^2} dt .
\end{align}

Meanwhile, the first term becomes
\begin{align}
  & \frac{1}{\tau^2} \int_0^\tau \int_0^\tau  \sum_{jk}\overline{ \delta a_j \delta a_k  \left( p_j X_j - p_j \delta f_j \right) \left( p_k  X_k - p_k \delta f_k \right) } dt dt'     \nonumber \\
  &\qquad\qquad=\int_0^\tau \int_0^\tau \sum_{jk} \delta a_j \delta a_k  p_j p_k \left(   -\gamma_j^2 p_j + \sum_l \gamma_l^2 p_l^2 \right) \left(  -\gamma_k^2 p_k + \sum_m \gamma_m^2 p_m^2 \right) dt dt'    \nonumber \\
& \qquad\qquad=  \int_0^\tau \int_0^\tau \left(  \left( \sum_j p_j^2 \gamma_j^2 \delta a_j \right) \left( \sum_k p_k^2 \gamma_k^2 \delta a_k \right) \right) dt dt' \nonumber \\
&\qquad \qquad= \left[ \int_0^\tau \left( \sum_j p_j^2 \gamma_j^2 \delta a_j \right) dt \right]^2,
\end{align}
where I used that $\sum_j p_j \delta a_j = 0$ in the third line.

 Combining the previous two equations gives
  \begin{align}
\overline{\left( \frac{1}{\tau} \int_0^\tau d \langle A \rangle - \langle dA \rangle - \cov(A,f)dt \right)^2} &= \frac{1}{\tau^2} \overline{\left(  \int_0^\tau    \sum_j p_j^2 \gamma_j^2 \delta a_j   dt \right)^2 } + \frac{1}{\tau^2} \overline{\int_0^\tau \sum_{j} \left(\delta a_j \right)^2 \gamma_j^2 p_j^2} dt \nonumber \\
& \leq \frac{1}{\tau^2} \overline{ \left(  \int_0^\tau   \sqrt{ \sum_j p_j^2 \gamma_j^4 } \sqrt{ \sum_k p_k^2 \left(\delta a_k\right)^2 }  dt \right)^2 } + \| \gamma \|_\infty^2 \frac{1}{\tau^2} \overline{\int_0^\tau  \sum_{j} \left(\delta a_j \right)^2 p_j^2 } dt \nonumber \\
& \leq \frac{1}{\tau^2} \overline{ \left(  \int_0^\tau   \sqrt{ \sum_j p_j \gamma_j^4 } \sqrt{ \sum_k p_k \left(\delta a_k\right)^2 }  dt \right)^2 } +  \| \gamma \|_\infty^2 \frac{1}{\tau^2} \overline{\int_0^\tau  \sum_{j} \left(\delta a_j \right)^2 p_j } dt \nonumber \\
&=  \overline{ \left( \frac{1}{\tau}\int_0^\tau \sigma_A  \, \sqrt{ \langle \gamma^4 \rangle } dt \right)^2 } + \frac{\| \gamma \|_\infty^2}{\tau} \overline{ \frac{1}{\tau} \int_0^\tau  \left( \sigma_A \right)^2 } dt.
\end{align}
The second line follows from the Cauchy-Schwarz inequality and from $\gamma_j \leq \| \gamma \|_\infty \coloneqq \max_j \{ \gamma_j \}$, and the third line holds because $p_j^2 \leq p_j$.

Note that the first term dominates the upper bound for long times $\tau \gg \| \gamma \|_\infty^2$. 

A looser bound holds by using that $\langle \gamma^4 \rangle \leq \|\gamma\|_\infty^4$,
\begin{align}
\overline{\left( \frac{1}{\tau} \int_0^\tau d \langle A \rangle - \langle dA \rangle - \cov(A,f)dt \right)^2} &\leq \| \gamma \|_\infty^4 \overline{\left( \frac{1}{\tau}\int_0^\tau \sigma_A \,   dt \right)^2 } + \frac{\| \gamma \|_\infty^2}{\tau} \overline{ \frac{1}{\tau} \int_0^\tau  \left( \sigma_A \right)^2 } dt.
\end{align}
This proves an equation analogous to Eq.~(14) in the main text for the stochastic replicator equation~(\ref{eq-app:stochreplicator}).
 
An analogous constraint holds when mutations are included, 
 \begin{align}
d p_j &= p_j \left( \sum_{k } p_k f_ k Q_{kj} - p_j \langle f \rangle -\gamma_j^2 p_j + \sum_l \gamma_l^2 p_l^2 \right) dt   + p_j \left( \gamma_j dW_j - \sum_l \gamma_l p_l dW_l \right), 
\end{align} 
in which case
\begin{align}
\overline{\left( \frac{1}{\tau} \int_0^\tau d \langle A \rangle - \langle dA \rangle - \sum_{jk} p_k (f_k - \langle f \rangle) Q_{kj} a_j dt \right)^2} &\leq \| \gamma \|_\infty^4 \overline{\left( \frac{1}{\tau}\int_0^\tau \sigma_A \,   dt \right)^2 } + \frac{\| \gamma \|_\infty^2}{\tau} \overline{ \frac{1}{\tau} \int_0^\tau  \left( \sigma_A \right)^2 } dt.
\end{align}

 \section{Rate limits under natural selection, mutations, and genetic drift}
\label{sec-app:ratelimitcomplete}
I prove Equation~(15) in the main text.

\vspace{7pt}

It holds that
\begin{align}
   \overline{ \left| \int_0^\tau \Big( d\langle A \rangle -  \langle dA \rangle \Big)\right| } - \overline{ \left| \int_0^\tau \sum_{jk} p_k Q_{kj} \big( f_k - \langle f \rangle  \big) a_j  dt \right| } &\leq \overline{ \left| \int_0^\tau   d\langle A \rangle -  \langle dA \rangle - \sum_{jk} p_k Q_{kj} \big( f_k - \langle f \rangle  \big) a_j  dt   \right| } \nonumber \\
   &\leq \sqrt{\overline{ \left| \int_0^\tau  d\langle A \rangle -  \langle dA \rangle  - \sum_{jk} p_k Q_{kj} \big( f_k - \langle f \rangle  \big) a_j  dt \right|^2 }} 
   \nonumber \\
   &\leq \| \gamma \|_\infty \sqrt{ \int_0^\tau \! \overline{ \left( \sigma_A \right)^2}  dt  }
\end{align}
where I used the triangle inequality in the first line, the Cauchy-Schwarz inequaltiy in the second line, and bound~\eqref{eq-app:auxRateLimitStochastic} in the third line.

The rate limit~(\ref{eq-app:AuxRateLimitMutations}), derived for the replicator-mutator equation, says that the replicator-mutator contribution to the change in $A$ satisfies
\begin{align}
\left| \sum_{jk} p_k Q_{kj} \big( f_ k - \langle f \rangle \big) a_j \right|  &= \left| \left[ \frac{d\langle A \rangle}{dt} - \langle \dot A \rangle \right] \bigg\vert_{\textnormal{replicator-mutator}} \right| \\
&\leq \langle f \rangle \big| \langle A \rangle_\Pi - \langle A \rangle \big| +    \sigma^\Pi_A \, \sigma_f.
\end{align}

Then, 
\begin{align}
    \overline{ \left| \int_0^\tau \Big( d\langle A \rangle -  \langle dA \rangle \Big)\right| }  \leq 
    \overline{\int_0^\tau\! \sigma^\Pi_A \, \sigma_f dt} 
   \, + \,     \overline{ \int_0^\tau \! \langle f \rangle \big| \langle A \rangle_\Pi - \langle A \rangle \big| dt} 
    \, + \, \| \gamma \|_\infty \sqrt{ \overline{ \int_0^\tau \!  \left( \sigma_A \right)^2  dt } }.
\end{align}
This proves Eq.~(15) in the main text.

\clearpage
\section{Mathematical toy models to illustrate the bounds}
\label{sec-app:example}

\begin{figure*}[ht]
  \centering  \framebox{\includegraphics[trim=00 00 00 00,width=0.951 \textwidth]{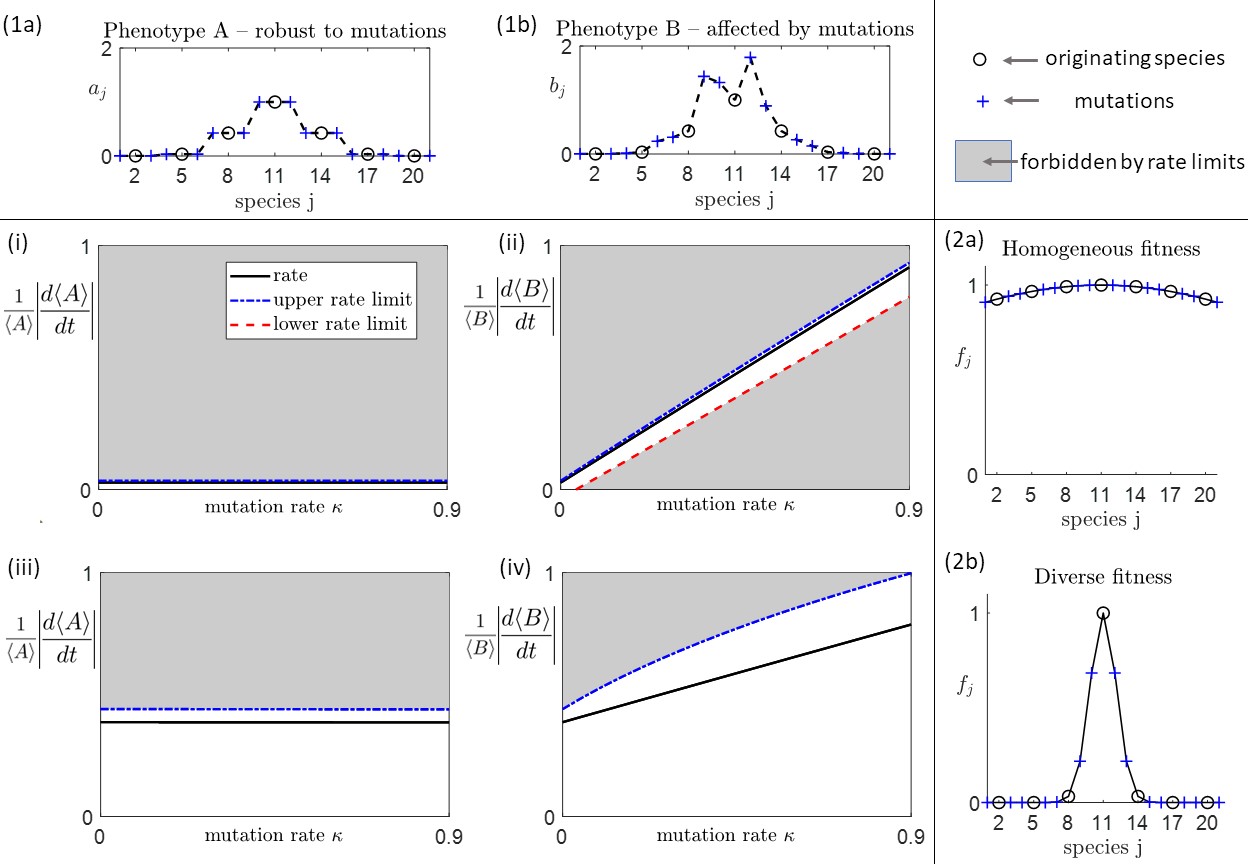}}\caption{\label{fig-app:fig1}\textbf{Evolutionary rate limits.}
I consider a toy model where $N = 7$ types, initially with equal populations,  mutate to two neighboring variants $j \longrightarrow \{j-1,j+1\}$ at rates $Q_{j,j-1} = 0.1 \kappa$ and $Q_{j,j+1} = \kappa$. While the replicator-mutator Eq.~(6) governs the evolution of the populations, the lower and upper rate limits Eq.~(9a) and~(9b) constrain the dynamics of the types' quantitative traits in terms of expectation values and standard deviations. For illustration purposes, I consider a phenotype $A$ (e.g., the flagella length of bacteria) whose values are not affected by the mutations [inset (1a)] and a phenotype $B$ (e.g., bacteria's mass) that changes on the mutated types [inset (1b)]. 
[Left column] For the mutation-robust phenotype $A$, it holds that $\langle A \rangle = \langle A \rangle_\Pi$, so  the upper rate limit Eq.~(9b) implies that $\left| d\langle A \rangle /dt \right| \leq \sigma^\Pi_A \, \sigma_f$.  Then, a homogeneous fitness landscape (weak-selection regime) for which $\sigma_f \ll 1$ [inset (2a)] results in slower changes in the phenotype [plot (i)] than the fast rates [plot (iii)] obtained with a diverse fitness profile (natural selection regime) for which $\sigma_f \gg 1$ [inset (2b)]. 
[Center column] The mutation-sensitive phenotype $B$ can evolve more rapidly than phenotype $A$ due to the contribution of the term $\langle f \rangle | \langle B \rangle_\Pi - \langle B \rangle |$ to the upper rate limit Eq.~(9b).  When the mutation-driven term $\langle f \rangle \left|\langle B \rangle_\Pi - \langle B \rangle \right|$ is larger than the natural selection contribution $\sigma^\Pi_A \sigma_f$, the lower rate limit Eq.~(9a) further constrains the minimum rates [plot (ii)].  The lower rate limits do not appear plots (i,iii,iv) since they are negative in those regimes. This example illustrates how the rate limits derived in this paper can be used to discriminate rapidly evolving traits from slowly evolving ones.}
\end{figure*}


\end{document}